\documentclass[published]{JHEP3}
\JHEP{08(2003)037}
 
\usepackage{amsmath, amsfonts,euscript,bbm}
\usepackage{epsfig}

\skip\footins = 1\bigskipamount plus 2pt minus 4pt

\newcommand{\UU}{\mathop{\rm {}U}\nolimits}
\newcommand{\SU}{\mathop{\rm SU}\nolimits}
 
\DeclareMathOperator{\F}{\EuScript{F}}
\newcommand{\mb}[1]{\boldsymbol{#1}}
\newcommand{\sD}[1]{\mbox{\bf D}_{#1}}
\newcommand{\sC}[1]{\mathcal Q_{#1}}
\newcommand{\Ket}[1]{\left|#1\right\rangle}
\newcommand{\nuBraKet}[2]{
   \llap{$\phantom{\Ket{#2}}$}_\nu\!\!\left.\left\langle#1\right|
   \! #2 \right\rangle_\nu}
\newcommand{\Expect}[1]{\left\langle #1 \right\rangle}
\newcommand{\NOrder}[1]{\text{:}#1\text{:}}
\DeclareMathOperator{\B}{B}
\def\hf{\frac12}
\def\ap{{\alpha^{\prime}}}
\def\Tr{\text{Tr}\;}
\def\DD{\text{D}\bar{\text{D}}}
\def\X{\mathcal{X}}
\def\Y{\mathcal{Y}}
\def\G{G}
\def\N{\mathcal N}
\def\nup{{\nu^\prime}}
\def\ints{\mathbbm{Z}}
\def\twist{\EuScript{T}}
\def\ie{i.e.~}
\def\sCoord{{\mb\theta}}
\def\BsCoord{{\mb\vartheta}}
\def\Tbar{{\bar T}}
\def\id{\mathbbm{1}}
\def\Zb{{Z^\dagger}}
\def\Psib{{\Psi^\dagger}}
\renewcommand{\Im}{\mathop{\rm Im}\nolimits}
\renewcommand{\Re}{\mathop{\rm Re}\nolimits}

\title{Spectral flow and boundary string field theory for angled
  D-branes}
 
\author{Nicholas T. Jones and S.-H. Henry Tye\\
        Laboratory for Elementary-Particle Physics, Cornell University\\
	Ithaca, NY 14853, USA\\
	E-mail: \email{nick.jones@cornell.edu},
	        \email{tye@mail.lns.cornell.edu}}

\abstract{D-branes intersecting at an arbitrary fixed angle
  generically constitute a configuration unstable toward
  recombination. The reconnection of the branes nucleates at the
  intersection point and involves a generalization of the process of
  brane decay of interest to non-perturbative string dynamics as well
  as cosmology.  After reviewing the string spectrum of systems of
  angled branes, we show that worldsheet twist superfields may be used
  in the context of Boundary Superstring Field Theory to describe the
  dynamics. Changing the angle between the branes is seen from the
  worldsheet as spectral flow with boundary insertions flowing from
  bosonic to fermionic operators.  We calculate the complete tachyon
  potential and the low energy effective action as a function of angle
  and find an expression that interpolates between the brane-antibrane
  and the Dirac-Born-Infeld actions.  The potential captures the
  mechanism of D-brane recombination and provides for interesting new
  physics for tachyon decay.}

\received{July 24, 2003}
\revised{August 18, 2003}
\accepted{August 22, 2003}

\begin{document}

\section{Introduction}\label{section1}

Properties of non-BPS configurations of D$p$-branes are very
interesting for a number of reasons. For one, these are unstable
dynamical systems and hence allow the study of time evolution for
non-perturbative objects in string theory.  With our now rather
elaborate understanding of the spectrum and the interactions of the
theory, a natural next step is then to study such dynamics. The
presence of tachyonic modes in non-BPS systems allows a number of new
scenarios: the tracking of tachyonic matter as it evolves in time, the
formation of defects (lower dimensional BPS D$p$-branes) as part of
the mechanism of tachyon condensation, the testing of the intricate
interconnections of closed string modes and open string dynamics.
Besides the recent interest in tachyon
rolling~\cite{Sen:2002nu,Sen:2002in} and
S-branes~\cite{Gutperle:2002ai,Strominger:2002pc} in string theory,
all these lend themselves to applications in cosmology as~well.

In brane inflation, an inflationary scenario in the brane world
describing the early universe, brane interactions and evolution
provide a natural origin to the inflaton and its
potential~\cite{Dvali:1998pa}--\cite{Blumenhagen:2002ua}.  A
particularly attractive scenario consists of a configuration of angled
branes~\cite{Garcia-Bellido:2001ky}.  The inflationary epoch of such a
system is relatively easy to study~\cite{Jones:2002cv} and is captured
by long distance physics. Toward the end of the inflationary epoch,
when the separation between branes becomes small compared to the
string scale, the lightest open string mode becomes tachyonic. As the
branes start intersecting, a rolling tachyon describes the
recombination of the branes, defect formation, and decay to closed and
open string modes. This is a fundamental interaction process in string
theory that is interesting in its own right: the evolutionary process
by which intersecting branes reconnect to minimize free energy.
Furthermore, the production of defects in brane inflation, which
appear as cosmic strings, can be tested in the near future in
cosmological observations~\cite{Sarangi:2002yt}. This means that a
stringy formulation of angled branes can be very useful for
phenomenology and can serve as a guide to properly embed string theory
into a cosmological scenario.  In this paper, we present a step in
this direction.

For parallel D$p$-branes in type IIA(B) theories with $p$ even (odd),
one typically has a BPS stable system; the massless open string modes
form a supersymmetric non-abelian gauge theory on the worldvolume,
often written at low energies as a Dirac-Born-Infeld (DBI) action. On
the other hand, for a non-BPS brane, or for a system of antiparallel
D$p$ branes known as a brane-antibrane pair, one has a
non-supersymmetric unstable configuration.  The relevant effective
actions~\cite{Kutasov:2000aq}--\cite{Jones:2002si} have been written
in the framework of Boundary Superstring Field Theory (BSFT)
~\cite{Witten:1992qy}--\cite{Marino:2001qc}.  Hence, to study the
regime interpolating between a configuration of parallel and
anti-parallel branes, it is natural to adopt BSFT techniques to write
a low energy effective action describing branes at an arbitrary
angle. As we change the angle between the two branes from 0 to $\pi$,
we would flow from a DBI action for two parallel D branes (DD system)
to the brane-antibrane action ($\mbox{D}\bar{\mbox{D}}$ system).
\FIGURE{\epsfig{file = Recombination.eps, width = 8cm,clip=}
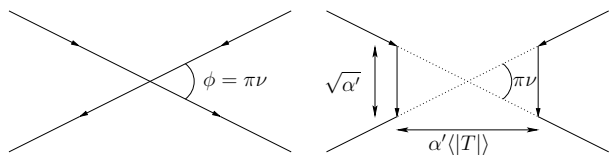
\caption{Angled branes and a crude description of their
recombination.\label{Rec3}}}
For generic non-zero angles, spacetime supersymmetry is completely
broken. For small separation between the bra\-nes, the lightest open
string mode is tachyonic and its mass runs as a function of angle.
The gradual rotation of the branes is seen from the string worldsheet
as spectral flow. As the angle increases from 0 to $\pi/2$, the
worldsheet fermion in the plane of the angle flows from an NS to a
Ramond fermion.  Further increasing the angle to $\pi$ makes the
fermion flow back to be NS, but with the opposite GSO projection than
at zero angle.  Since the tachyonic mode is our main focus, we shall
measure the angle $\phi$, or $\nu\equiv\phi/\pi$, with respect to the
brane-antibrane system; that is, $\nu = 0$ corresponds to the $\DD$
configuration, while $\nu = 1$ corresponds to the supersymmetric DD
system.

For $0 \le \nu \le 1$, we are then dealing with worldsheet physics
reminiscent of strings on orbifolds. The brane angle plays the role of
the orbifold twist yielding worldsheet bosons and fermions with
non-integer moding. In the case at hand, the twist $\nu$ can be an
arbitrary irrational number.  This introduces the language of twist
fields in the BSFT technology requiring a careful treatment of the
monodromy properties of worldsheet operators. Among other novelties,
the presence of the twist fields implies that what used to be a free
and explicitly calculable worldsheet theory becomes interacting, so
only a perturbative expansion can be performed for some quantities.
As a result of this complication, we can determine the effective
action from BSFT with twists only up to the canonical kinetic term,
but the tachyon potential can be obtained exactly.

Our main results can be summarized with the following expression for
the low energy effective action of the angled D8-branes system.
\begin{eqnarray}
S(T,\Tbar) &=& \int d^8x\; V_\nu(T)\sqrt{-\text{P}[g]} \left( 1 +
\mathcal K_\nu\;\partial_aT \partial^a\Tbar \right) + \cdots
\nonumber\\
V_\nu(T) &=& N_\nu \prod_{n=0}^\infty \frac
{\displaystyle\sin\left(\frac{\sin(\tfrac{\pi\nu}2) 2\pi\ap T\Tbar}
{[n+\hf(\nu+1)]\B(n+1,\nu)} + \frac{\pi\nu}2\right)}
{\displaystyle\sin\left(\frac{\sin(\tfrac{\pi\nu}2)}{n\B(n,\nu)}
  2\pi\ap T\Tbar + \frac{\pi\nu}2\right)},
\nonumber\\ 
K_\nu &=& 2\pi\ap^2\frac{\sqrt\pi\Gamma(1-\frac\nu2)}
{2^\nu\Gamma(\frac{3-\nu}2)}\left\{1 + (1-\nu) \left[\Psi(2-\nu)
  -\Psi\left(\frac{3-\nu}2\right)\right]\right\},
\nonumber
\end{eqnarray}
in which $\B(p,q) \equiv \Gamma(p)\Gamma(q) / \Gamma(p+q)$ and
$\Psi(y) \equiv \tfrac{d}{dy}\ln\Gamma(y)$ are the Beta and Polygamma
functions respectively, P$[g]$ is the pull-back of the metric to the
brane world-volume, and `$a$' runs over spacetime directions
transverse to the plane of intersection.  We argue on physical grounds
that the potential is normalized to be
\begin{equation}
N_\nu = 2\tau_8 \sqrt{\frac{2\pi\ap}{\cos(\tfrac{\pi\nu}2)}}\,,\qquad
V_\nu(0) = 2\tau_8
\sqrt{\frac{2\pi\ap}{\cos(\tfrac{\pi\nu}2)\sin(\tfrac{\pi\nu}2)}}\,.
\end{equation}
Furthermore, some dependence on worldvolume gauge fields can be
introduced in this action by covariantizing the spacetime derivatives.

These complicated expressions have the correct properties to
interpolate between the unstable $\DD$ system and the BPS DD system.
For small $\nu$, the behaviour of the potential is well approximated
by the first term in the infinite product.  For $\nu < 1$, the complex
field $T$ is tachyonic, and will roll to an expectation value of
$\sqrt{2\pi\ap}\Expect{|T|} \sim \sqrt{\cot (\tfrac{\pi\nu}2)}$, at
which point the potential achieves its minimal value of
$V_\nu(\Expect{|T|})/V_\nu(0) \sim \sqrt{\sin (\tfrac{\pi\nu}2)}$.
Note that the minimum moves from $T\to\infty$ when $\nu=0$ to $T=0$
when $\nu=1$, agreeing with the expected physics of the system.
Beyond the stable minimum of the potential, the potential increases
and diverges at $2\ap T\Tbar = (1-\nu/2) / \sin{\pi\nu}/2$.
The kinetic term normalization ensures that the tachyon mass flows
smoothly with $\nu$ from that of the BSFT $\DD$ tachyon to $0$ for the
stable DD system. Dynamics in this potential involves recombination of
the branes (see figure~\ref{Rec3}), followed by the decay of the tachyon
condensate - after the roll to the minimum of the potential where
several channels are available for dumping energy into open string
modes; for example, the gauge fields transverse to the angling plane
incorporated in the low energy effective action by covariantizing the
derivatives.

 This paper is organized as follows: section~\ref{section2} is a brief
 review of the spectrum of strings for the angled D-brane system.  In
 section~\ref{section3}, we describe how the language of twist fields
 is a convenient tool to study angled branes, and we develop the
 boundary twist field formalism to apply to the BSFT of branes at
 angles.  We find that the twist fields flow from being bosonic to
 fermionic boundary fields through worldsheet spectral flow.
 Section~\ref{section4} is a review of the BSFT for the
 brane-antibrane system, and in section~\ref{section5} we perform BSFT
 calculations for angled branes, and derive the tachyon low energy
 effective action.  In section~\ref{section6}, we illustrate that our
 results give an effective description of the recombination process.
 We discuss the results and future work in section~\ref{section7}.
 \textref{section8}{Appendix} is a summary of our superspace and CFT
 conventions.

\section{Spectrum of angled branes}\label{SecSpec}\label{section2}

In this section, we review the spectrum of strings stretching between
two D-branes of the same dimension intersecting at an arbitrary
angle~\cite{Berkooz:1996km}--\cite{Polchinski:1998rr}.  For simplicity
we work with two D8-branes in IIA theory angled in the 8-9 plane.  In
this section, the worldsheet is represented by the strip parameterized
by $(\tau,\sigma)$ with boundaries located at $\sigma = 0,\pi$.  We
introduce a complex structure in the 8-9 directions with new
coordinates $Z = 2^{-1/2}[X^8(z,\bar z) + i X^9(z,\bar z)]$, $\Psi(z)
= 2^{-1/2}[\psi^8(z) + i \psi^9(z)]$, where the $\psi$'s are
worldsheet fermions in the NSR formalism. With one brane lying along
the $X^8$ direction and the other tilted by an angle $\phi \equiv
\nu\pi$, the boundary conditions on $Z$ for a string stretching
between the branes are
\begin{eqnarray}
   &\mbox{at}\ \ \sigma = 0\quad
   \begin{cases}
     \partial_\sigma \Re Z = 0\,,\\
     \Im Z = 0\,,
   \end{cases}
   &\ \ \mbox{at}\ \ \ \sigma = \pi\quad
   \begin{cases}
     \partial_\sigma \Re\left(e^{i\pi\nu}Z\right) = 0\,,\\
     \Im \left(e^{i\pi\nu}Z\right) = 0\,,
   \end{cases}
\nonumber
\end{eqnarray}
and that for the worldsheet fermion $\psi$ is
$$
   \Psi(\tau,0) = e^{-2\pi i\nup}\Psi(\tau,2\pi)\,.
$$ 
We have used the standard doubling trick to extend the fermions on
$\sigma = [0,\pi]$ to $\sigma = [0,2\pi)$.  Note that $0\leq\nu\leq
  1$, with the configuration corresponding to $\nu=0$ being
  conventionally the brane-antibrane system and $\nu=1$ corresponding
  to the brane-brane system.  The twist $\nup$ is $\nu + 1/2$ for NS
  sector fermions and $\nu$ for Ramond sector fermions.

Transforming to the upper half plane (UHP) through $z =
e^{\sigma+i\tau}$, with the boundary now being along the real axis,
the mode expansions for the free string become
\begin{equation}
\begin{array}[b]{rclcrcl}
Z(z,\bar z) &=&\displaystyle i\left(\frac{\ap}2\right)^{1/2}
   \sum_{n\in\ints} \left[
     \frac{\alpha^\dagger_{n-\nu}}{(n-\nu)z^{n-\nu}}
     +\frac{\alpha_{n+\nu}}{(n+\nu)\bar z^{n+\nu}}\right],
   &\qquad& \Psi(z) &=& \displaystyle\sum_{r\in\ints-\nup}
   \frac{\psi_r}{z^{r+1/2}}\,,\\
   \Zb(z,\bar z) &=&\displaystyle i\left(\frac{\ap}2\right)^{1/2}
   \sum_{n\in\ints} \left[
     \frac{\alpha_{n+\nu}}{(n+\nu)z^{n+\nu}}
     +\frac{\alpha^\dagger_{n-\nu}}{(n-\nu)\bar z^{n-\nu}}\right],
   &\qquad& \Psib(z) &=&\displaystyle \sum_{s\in\ints+\nup}
   \frac{\psi^\dagger_s}{z^{s+1/2}}\,.\label{Modes}
 \end{array}
\end{equation}
For these expansions, the negative real axis has one boundary
condition, and the positive real axis the other; hence there is a flip
in boundary conditions at $z = 0$ and $z = \infty$.  Quantization of
the string leads to the oscillator algebra
$$
\bigl[\alpha_{m+\nu},\alpha^\dagger_{n-\nu}\bigr] =
(m+\nu)\delta_{n+m=0}\,,\qquad \bigl\{\psi_r,\psi^\dagger_s\bigr\} =
\delta_{r+s=0}\,.
$$

To compute the spectrum, we note the relations
\begin{equation}
\begin{array}[b]{rclcrcl}
\bigl[\alpha^\dagger_{n-\nu},L_0\bigr] 
&=& (n-\nu)\alpha^\dagger_{n-\nu}\,,&\qquad&
   \bigl[\alpha_{n+\nu},L_0\bigr] &=& (n+\nu)\alpha_{n+\nu}\,,
\\
   \bigl[\psi_r,L_0\bigr] &=& r\psi_r,
   &\qquad& \bigl[\psi^\dagger_s,L_0\bigr] &=& s\psi^\dagger_s\,,
\end{array}
\end{equation} 
with $L_0 = (\sum_{n\in\ints} :\alpha^\dagger_{n-\nu}\alpha_{-n+\nu}:
+ \sum_{r\in\ints-\nup}r:\psi^\dagger_{-r}\psi_{r}: ) + E_0$, where
the total zero point energy is
$$
E_0 = \hf(\nu - 1)\quad\text{in the NS sector};
E_0 = 0\quad\text{in the R sector}\,.
$$
We then define the twisted vacuum ${\Ket0}_\nu$ by
 \begin{equation}
\begin{array}[b]{rclcrcl}
\alpha^\dagger_{n-\nu}{\Ket0}_\nu &=& 0,\quad n\ge1\,,
   &\qquad& \psi_{n-\nup}{\Ket0}_\nu &=& 0,\quad n\ge1\,,
\\
   \alpha_{m+\nu}{\Ket0}_\nu &=& 0\,,\quad m\ge0\,,
   &\qquad& \psi^\dagger_{m+\nup}{\Ket0}_\nu &=& 0\,,\quad m\ge0\,.
 \end{array}
\end{equation}
Note that flowing from $\nu=0$ through $\nu=1/2$, $\Ket0_\nu$ becomes
the first excited state in the NS sector, and
$\psi_{-\hf+\nu}\Ket0_\nu$ becomes the state of lowest energy.

\DOUBLEFIGURE[t]{NSFlow.eps}{BosonicFlow.eps}
	 {\label{NSFlow}
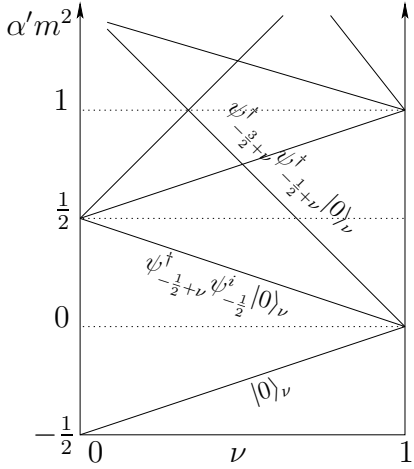
	   The NS$+$ spectrum of the states stretched
	   between a brane pair at an angle $\phi = \pi\nu$.  From the
	   brane-antibrane system $\nu=0$, the spectrum flows to the
	   NS$+$ sector for the brane-brane system at $\nu=1$.
	   $\psi^i_{-1/2}$ are the fermionic raising operators in the
	   transverse directions in which the branes are parallel.
	 }
	 {\label{BoseFlow}
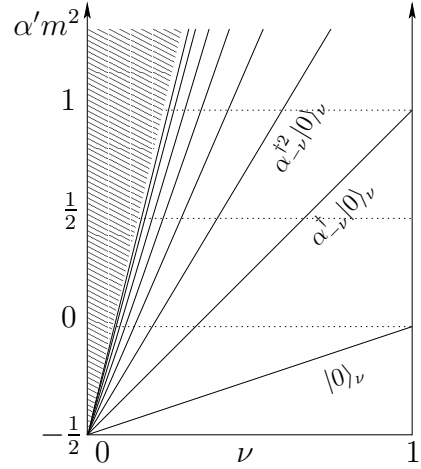
	   The bosonic excitations of the tachyonic NS vacuum which
	   become momentum modes when $\nu=0$; the states of different
	   masses correspond to KK-like excitations of the lowest
	   state, and the spectrum is indicative of an expansion in
	   states localized about the intersection. $\Ket0_\nu$ is the lowest
	   lying tachyon that we focus on in this work.
	 }

We consider the spectrum starting with the brane-antibrane (NS$-$)
spectrum at $\phi = \pi\nu = 0$ which flows to the brane brane (NS$+$)
spectrum at $\phi = \pi\nu = \pi$.  For $\nu>0$ there is a splitting
of the spectrum since the $\psi$ oscillators in the different
directions have different weights. The NS spectrum can be easily
computed from the results above and is described in
figure~\ref{NSFlow}.

It is important to note also that the bosonic oscillators,
$\alpha^\dagger_{-\nu}$ and $\alpha_{-1+\nu}$ are momentum oscillators
at $\nu = 0$ and $\nu = 1$ respectively.  Close to the brane-antibrane
($\nu = 0$) the modes created by $\alpha^\dagger_{-\nu}$ acting on the
lowest tachyonic mode can still be tachyonic, as shown in
figure~\ref{BoseFlow}.  As the angle is decreased, additional
tachyonic states appear at discrete intervals of the angle given by
$\nu=1/(2 n+1)$ with $n=0,1,2,\ldots$ For $\nu\neq 0$ or $1$ these
states are localized near the intersection point of the branes.  Their
masses decrease with $\nu$ as shown in figure~\ref{BoseFlow} until
they all become degenerate with the lowest lying tachyon at
$\nu=0$. Hence, we have an infinite tower of tachyonic states
collapsing on top of each other at $\nu=0$, reminiscent of
decompactification in Kaluza Klein theories.  In this case, we can
think of the process as decompactification along the brane
worldvolume: when the branes become antiparallel at $\nu=0$, the
tachyon is no more localized at an intersection point and can travel
along the worldvolume.  As we will see, the localization of this tower
of tachyonic states helps in smoothing out the shape of the bent
branes at their intersection point as they undergo recombination.
Finally, note also that a mirror picture takes hold on the other side
at $\nu=1$. In that region, a infinite number of massive states
collapse to zero mass comprising the modes of (gauge) fields localized
at the intersection point.

\section{Boundary twist fields}\label{BTwist}\label{section3}

The operator formalism used to compute the spectrum in the previous
section can be used to compute correlators in the angled brane system
when one can arrange for the boundary condition flips to occur at
$z=0$ and $z=\infty$ along the boundary.  In the BSFT formalism, it
will be necessary to consider a condensate of perturbations located at
arbitrary points. The operator formalism hence quickly becomes
cumbersome and it is more convenient to adopt a different approach.

The problem of angled branes can be recast in the language of twist
fields that incorporate the change in the moding of the worldsheet
fields when acting upon twisted vacua.  Such twist operators must be
present when boundary conditions flip locally across insertion points
to move from one endpoint of the open string located on one brane to
the other~\cite{Cardy:1989ir}.  We will first review the use of such
operators as they have risen in the past in the context of studying
strings on orbifolds.  We will then move onto applying the technology
to the system of angled branes and BSFT.

\subsection{Review of twist fields}\label{section3.1}

The expectation value of a stress energy tensor in a twisted vacuum is
non-vanishing~\cite{Dixon:1987qv} and can be easily computed
$$
   \left\langle T^Z_B(z) \right\rangle 
   = \frac{\hf\nu(1-\nu)}{z^2}\nuBraKet{0}{0}\,,\qquad
   \left\langle T^\Psi_B(z) \right\rangle 
   = \frac{\hf(\nup-\hf)^2}{z^2}\nuBraKet{0}{0}\,.
$$ 
The double poles at $z=0$ suggest that there is a local source of
stress-energy at $z=0$ whose origin is a local operator.  These new
operators of conformal dimension $h = 1/2\nu$ in the NS sector (with
$\nup = 1/2 + \nu$) or $h=1/8$ in the R sector (with $\nup = \nu$)
factor the information about the twist away from the vacuum.  One then
defines bosonic twist operators $\sigma^\pm$ having the OPEs (on the
UHP)~\cite{Dixon:1987qv}\footnote{Because we are working with open
  strings, the holomorphic and anti-holomorphic excited twist fields
  are identified, whereas for the orbifold twist fields of
  ~\cite{Dixon:1987qv}, they are distinct.}
\begin{equation}
\begin{array}[b]{rclcrcl}
   \partial Z(z)\sigma^+(0) &=& \displaystyle\frac1{z^{1-\nu}}\mu^+(0)\,,
   &\qquad& \bar\partial Z(\bar z)\sigma^+(0) &=& 
   \displaystyle\;\frac1{{\bar z}^\nu}\;\mu^{\prime +}(0)\,,
\\[7pt]
   \partial \Zb(z)\sigma^+(0) &=&\displaystyle 
   \;\frac1{z^\nu}\;\mu^{\prime +}(0)\,,
   &\qquad& \bar\partial \Zb(\bar z)\sigma^+(0) &=&\displaystyle 
   \frac1{{\bar z}^{1-\nu}}\mu^+(0)\,,
\\[7pt]
   \partial Z(z)\sigma^-(0) &=&\displaystyle\frac1{z^\nu}\;\mu^{\prime -}(0)\,,
   &\qquad& \bar\partial Z(\bar z)\sigma^-(0) &=&\displaystyle 
   \frac1{{\bar z}^{1-\nu}}\mu^-(0)\,,
\\[7pt]
   \partial \Zb(z)\sigma^-(0) &=&\displaystyle 
   \frac1{z^{1-\nu}}\mu^-(0)\,,
   &\qquad& \bar\partial \Zb(\bar z)\sigma^-(0) &=&\displaystyle 
   \frac1{{\bar z}^\nu}\;\mu^{\prime -}(0)\,.
\label{BTwistOPEs} 
\end{array}
\end{equation}
These OPEs are determined as in~\cite{Dixon:1987qv} (where $\nu$ was a
rational number) by the requirements that $Z$ acquires a phase of
$e^{\pm i\nu\pi}$ across the change in boundary condition implemented
by $\sigma^\pm$; that the bosonic derivative fields create excited
states; and that $\sigma^\pm$ are to be fields of highest weight.
Since in this analysis all twist fields are to be placed on the
worldsheet boundary, these relations hold for any $\nu \in [0,1]$,
rational or otherwise, since we will never encounter ambiguous contour
integrals involving branch cuts on the $z$-plane. Hence, the conformal
weights of these twist and excited twist fields are
$$
   h_{\sigma^\pm} = \frac\nu2(1-\nu)\,,\qquad
   h_{\mu^\pm} = \frac\nu2(3-\nu)\,,\qquad
   h_{\mu^{\prime \pm}} = 1 - \frac\nu2 - \frac{\nu}2^2\,.
$$

Similarly, there are operators $s^\pm$ which twist the worldsheet
fermions which have the OPEs
\begin{equation}
\begin{array}[b]{rclcrcl}
\Psi(z) s^+(0) &=&\displaystyle z^{\nup-1/2} u^{\prime +}(0)\,,&\qquad&
   \tilde\Psi(\bar z)s^+(0) 
   &=&\displaystyle \frac1{{\bar z}^{\nup-1/2}}u^+(0)\,,
\\[7pt]
   \Psib(z) s^+(0) &=&\displaystyle \frac1{z^{\nup-1/2}}u^+(0)\,,
   &\qquad& \tilde\Psi^\dagger(\bar z)s^+(0) 
   &=&\displaystyle {\bar z}^{\nup-1/2}u^{\prime +}(0)\,,
\\[7pt]
\Psi(z) s^-(0) &=&\displaystyle \frac1{z^{\nup-1/2}}u^-(0)\,,
   &\qquad& \tilde\Psi(\bar z)s^-(0) 
   &=& \displaystyle{\bar z}^{\nup-1/2}u^{\prime -}(0)\,,
\\[7pt]
\label{FTwistOPEs}
   \Psib(z) s^-(0) &=&\displaystyle z^{\nup-1/2} u^{\prime -}(0)\,,
   &\qquad&\tilde\Psi^\dagger(\bar z) s^-(0) 
   &=& \displaystyle\frac1{{\bar z}^{\nup-1/2}}u^-(0)\,.
\end{array}
\end{equation}
 These are again determined by requiring that $\Psi$ picks up the
 appropriate phase, with the ``$+$'' and ``$-$'' fields twisting the
 fermions in opposite directions.  The OPEs are given for both NS
 ($\nup = \nu + 1/2$) and R ($\nup = \nu$) fermions. Henceforth, we
 shall focus only on the NS sector and set $\nup = \nu + 1/2$ as this
 entails the interesting dynamics of branes at angles.  Note that by
 bosonizing the fermions, we may represent the fermionic sector in
 bosonic variables in the standard way
\begin{equation}
\begin{array}[b]{rclcrclcrcl}
\Psi(z) &\sim& e^{iH(z)}\,,&\qquad&
   \Psib(z) &\sim& e^{-iH(z)}\,,&\qquad& &{}&
   \\
   s^\pm(z) &\sim& e^{\pm i \nu H(z)}\,,&\qquad&
   u^\pm(z) &\sim& e^{\mp i(1-\nu) H(z)}\,,&\qquad&
   u^{\prime \pm}(z) &\sim& e^{\pm i (1+\nu)H(z)}\,,
\label{Bosonise}
 \end{array}
\end{equation}
where $H$ is a holomorphic bosonic field.  The anti-holomorphic side
has similar bosonization.  The conformal weights of the twist fields
$s$ and $u$ are then given by
$$
   h_{s^\pm} = \frac{\nu}2^2\,,\qquad
   h_{u^\pm} = \hf(1-\nu)^2\,,\qquad
   h_{u^{\prime \pm}} = \hf(1+\nu)^2\,.
$$

To construct the boundary condition changing operator, we require a
field which twists both the bosons and the fermions.  In
~\cite{Dixon:1987qv}, it is argued that in a supersymmetric theory,
the twist on $\Psi$ must compensate the twist on $Z$ in order that the
worldsheet supercurrent $T_F$ is single or double valued in the NS and
R sectors respectively.  The operator of lowest dimension which
accomplishes this is $\twist_0^\pm \equiv(\sigma s)^\pm$ (although
other combinations of twist fields which satisfy this condition are
also important in this system; see section~\ref{Vertices} for an
example).  These twist fields have a non-trivial supersymmetry
transformation so they can be written as the lower component of a
twist superfield.  Since we work on a worldsheet with a boundary and
the twist operators are to be placed only on the boundary, they must
represent $\N = 1$ supersymmetry rather than the $\N = (1,1)$
supersymmetry of the worldsheet bulk.  Hence, $\N=1$ superfields will
be used as BSFT boundary insertions that preserve the boundary
supersymmetry while breaking conformal invariance.

Using the conventions of~\cite{Polchinski:1998rr} (which differ from
those in~\cite{Dixon:1987qv}), our twist superfield becomes
\begin{equation}
\label{BTwistSuF}
\mb\twist^\pm \equiv \twist_0^\pm + i\BsCoord\twist_1^\pm \equiv
(\sigma s)^\pm - \frac i{\sqrt{2\ap}}\BsCoord (\mu u)^\pm\,.
\end{equation}
See \textref{section8}{appendix} for details on deriving this expression.
The top component of the twist superfield $\twist_1$ has conformal
weight $h = \hf(\nu+1)$.  This is essentially the holomorphic side of
the twist superfield constructed in~\cite{Dixon:1987qv}, which in the
present case is linked to the anti-holomorphic side through the
boundary conditions.

\subsection{Spectral flow}\label{section3.2}

In inserting perturbations on the open string worldsheet boundary, it
is convenient to represent the worldsheet by the unit disk.  We will
then consider the worldsheet as being defined by $|z|<1$ with boundary
$|z|=1$.  We parametrize this disk by $z=e^{-i \tau +\sigma}$ such
that $\tau$ is an angular coordinate along the boundary.  In the
correspondences to follow we map the fields to the boundary of the
unit disk with this coordinate $\tau$, so that holomorphic and
anti-holomorphic fermions are linked by the boundary conditions as
described in \textref{SuAppendix}{appendix}, and the derivatives are
restricted to $\tau$ derivatives along the boundary.

It is evident from the OPEs~(\ref{FTwistOPEs}), and particularly the
bosonization~(\ref{Bosonise}), that in the limiting cases of no twist
$\nu=0$ and a full twist $\nu=1$ the fermionic twist fields flow as
 \begin{align*}
   &&\nu =& 0 & \longrightarrow && \nu =& 1\\
   s^+, s^-:&& \id && \longrightarrow && \Psi,& \Psib\\
   u^+,u^-:&& \Psib,& \Psi &\longrightarrow && \id&\\
   u^{\prime +}, u^{\prime -}:&& \Psi,\;& \Psib & \longrightarrow && 
   \NOrder{\dot\Psi\Psi},\;&\NOrder{\dot\Psi^\dagger\Psib}
 \end{align*}

Similarly, the limiting cases of the bosonic twist fields can be
deduced from the OPEs; this is more subtle without the aid of a free
field description for the twist fields, but the OPEs are sufficient to
identify the endpoints of the flow.  Although the resultant
identifications are not unique the general structure can be obtained.
When $\nu\to0$, there remains a singularity in the OPE of $\partial Z$
with $\sigma^+$. This identifies $\sigma^+$ with some combination of
$\NOrder{(\Zb)^n}$ as the only structure to have a non-trivial simple
pole in the OPE with $\partial Z$.  Furthermore, since the operator
multiplying that pole is just $n\NOrder{(\Zb)^{n-1}}$, $\mu^+$ is also
identified in the limit $\nu\to0$.  We then propose the mappings
\begin{align*}
   &&\nu = &0 &\longrightarrow&& \nu =& 1\\
   \sigma^+:&& \NOrder{(\Zb&)^n} &\longrightarrow&& \NOrder{(Z&)^n}\\
   \mu^+:&& -(2\ap)n\NOrder{(\Zb&)^{n-1}} 
   &\longrightarrow&& \NOrder{\partial Z(Z&)^n}\\
   \mu^{\prime +}: && \NOrder{\partial Z^\dagger(\Zb&)^n} 
   &\longrightarrow&& -(2\ap)n\NOrder{(Z&)^{n-1}}\,,
 \end{align*}
where $n\ge0$.  The relations satisfied by the ``$-$'' twist fields
are obtained by swapping $Z$ and $\Zb$ in the table above. Note that
for brevity we have now adopted a new notation for the $\partial$
symbol: $\partial Z \equiv 2^{-1/2}(\dot X^8+i(X^9)^\prime)$, where
the prime is the normal derivative to the boundary.  The other
derivatives are zero as dictated by the boundary conditions.  The UHP
boundary conditions on the fields are actually encoded within the
OPEs~(\ref{BTwistOPEs}), since their $\nu=0\ (1)$ cases identify
$\partial\Zb(\partial Z)$ with $\bar\partial Z(\bar\partial\Zb)$.  The
$n=0$ case of the relations above is then
\begin{align*}
   &&\nu =& 0 &\longrightarrow&& \nu =& 1\\
   \sigma^\pm:&& \id &&\longrightarrow&& \id&\\
   \mu^+, \mu^-:&& 0 &&\longrightarrow&& 
   \partial Z, &\partial Z^\dagger\\ 
   \mu^{\prime +}, \mu^{\prime -}: 
   && \partial Z^\dagger,& \partial Z &\longrightarrow&& 0\,.&\\
\end{align*}
We argue that the $n=0$ case is the correct limiting behaviour for the
bosonic twist fields; if the twist fields $\sigma^\pm$ are to
represent the boundary condition changing operators, then in the
no-twist and full-twist cases they must reduce to the identity
operator, since the boundary conditions on both branes are identical.
In the next section we construct vertex operators under this
assumption, and see that they exactly reproduce the tachyon and gauge
field vertex operators in the corresponding limits.  The formalism to
follow is consistent irrespective of whichever $n$ is chosen, and the
flow of the fermions established from the bosonization guides the
structure.

Combining the fermion and the $n=0$ boson limiting expressions, we
obtain the limiting expressions on the twist superfields:
 \begin{align*}
   &&\nu = 0 &&\longrightarrow && \nu &= 1\\
   \mb\twist^+ &= \twist_0^+ + i\BsCoord\twist_1^+:&
   \id + \BsCoord0 &&\longrightarrow&&
   -\frac i{\sqrt{2\ap}}(i\sqrt{2\ap}&\Psi + \BsCoord\partial Z)\\
   \mb\twist^- &= \twist_0^- + i\BsCoord\twist_1^-:&
   \id + \BsCoord0 &&\longrightarrow&&
   -\frac i{\sqrt{2\ap}}(i\sqrt{2\ap}&\Psib + \BsCoord\partial \Zb).
 \end{align*}
These assignments agree with all the OPEs, and they will be shown to
enable the BSFT reconstruction of the flow of the states in the
spectrum from GSO odd to GSO even states described in
figures~\ref{NSFlow} and \ref{BoseFlow}.

As noted earlier, the GSO$-$ vacuum for $\nu<1/2$ becomes the GSO$+$
first excited state for $\nu>1/2$, which is generated by a fermionic
excitation of the ground state.  This is the spectral flow described
in~\cite{Lerche:1989uy}, and shall \pagebreak[3] be manifest on the worldsheet
through the flow of the twist fields from bosonic to fermionic
statistics.  Thus the tachyon will become part of the $\UU(2)$ gauge
field which is the string state excited by a fermionic oscillator from
the vacuum, in the off-diagonal Chan-Paton sector.

\subsection{Twist field correlators}\label{TFCorrelators}\label{section3.3}

In the BSFT of angled branes, since one inserts the twist operators we
encountered in the previous section on the boundary of the worldsheet
at $|z|=1$, correlators of the twist operators in the free conformal
theory are needed.  We can write the correlator
\begin{equation}
\label{Twist0Correlator}
\Expect{\twist_0^-(z_1)\twist_0^+(z_2)} = \frac{1}{(z_1 - z_2)^\nu}\,.
\end{equation}
The power of $(z_1-z_2)$ is fixed by the conformal weight of the
$\twist_0^\pm$ operators (each being $1/2\nu$), and the coefficient is
fixed by the fact that in the $\nu\to$ 0 and 1 limits, the trivial
correlator of $\id$ and the fermion propagator are to be reproduced.
In principle we can multiply this correlator by any non-singular
function of $\nu$ which is 1 when $\nu = 0$ and $1$, but any such
additional $\nu$ dependence can be removed by rescaling the twist
fields. As written~(\ref{Twist0Correlator}) flows smoothly between
being trivial for $\nu = 0$, to the correlator of a free fermion at
$\nu = 1$.

Changing coordinates to $\tau$, the angular coordinate along the disk
boundary, we then have
 \begin{eqnarray}
   \Expect{\twist_0^-(\tau_1)\twist_0^+(\tau_2)} &=& 
   \frac{(-iz_1)^{\hf\nu}(-iz_2)^{\hf\nu}}{(z_1 - z_2)^\nu}
   \\
   &=& \frac1{\left[2\sin\left(\frac{\tau_1-\tau_2}2\right)\right]^\nu}\,.
 \end{eqnarray}
The correlator we computed naturally incorporates time ordering as
necessitated by a path integral formulation. When decomposing this
correlator in frequency modes, unlike correlators of half integer
dimension operators there are two branch choices and we need to be
careful to pick branches consistent with the time ordering of
$\twist_0^-$ and $\twist_0^+$.  With $\epsilon \in (0,2\pi)$, we
identify two possible cases:
\begin{eqnarray}
   \Expect{\twist_0^-(\epsilon)\twist_0^+(0)} &=& e^{-i\tfrac\pi2\nu} 
   \sum_{n=0}^\infty \frac{e^{i\epsilon(n+\hf\nu)}}{n\B(n,\nu)}\,,
\nonumber\\
   \Expect{\twist_0^-(2\pi-\epsilon)\twist_0^+(0)} &=& 
   \Expect{\twist_0^+(0)\twist_0^-(-\epsilon)} = 
   \Expect{\twist_0^+(\epsilon)\twist_0^-(0)} =
   e^{i\tfrac\pi2\nu} \sum_{n=0}^\infty 
   \frac{e^{-i\epsilon(n+\hf\nu)}}{n\B(n,\nu)}\,,
\end{eqnarray}
where $\B(p,q)$ is the Beta function.  We have associated positive
frequency modes with $\twist_0^-$ being ahead of $\twist_0^+$, and
negative frequency modes for the other ordering. In this context, it
is being assumed that the expansions are made convergent by adding a
small imaginary part to $\epsilon$ as appropriate for each of the two
branches. This representation will be useful later when we will need
the correlators in the interacting theory.  We then have\footnote{ It
  is worthwhile noting that this issue is slightly more subtle in the
  case at hand, as opposed to the special situation when $\nu=1$
  addressed in the literature. The subtlety arises as frequency modes
  $n+(\nu/2)$ with $n\in\{-\infty,\infty\}$ are different from
  $-(n+(\nu/2))$ when $\nu$ is not zero or one.}
$$
   \Expect{\twist_0^\mp(-\epsilon)\twist_0^\pm(0)} 
   = e^{\mp i\pi\nu} \Expect{\twist_0^\mp(\epsilon)\twist_0^\pm(0)}\,.
$$ The phase can be understood in terms of the phase some quantities
acquire under a change in boundary conditions.

Calculating the correlator on the disk boundary of the excited twist
operators in the same way, we get
 \begin{eqnarray}
\nonumber
   \Expect{\twist_1^-(z_1)\twist_1^+(z_2)}
   &=& \frac{\nu}{(z_1-z_2)^{\nu+1}},
   \\
\nonumber
   \Expect{\twist_1^-(\tau_1)\twist_1^+(\tau_2)} &=& \nu
   \frac{(-iz_1)^{\hf(\nu+1)}(-iz_2)^{\hf(\nu+1)}}{(z_1-z_2)^{\nu+1}}
   \\
   &=&
   \frac\nu{\left[2\sin\left(\frac{\tau_1-\tau_2}2\right)\right]^{\nu+1}}\,. 
\label{Twist1Correlator}
 \end{eqnarray}
As $\nu:0\to1$ this correlator smoothly flows from zero to
$\tfrac1{2\ap}\Expect{\partial Z \partial \Zb}$.  The normalization of
this correlator was also calculated in~\cite{David:2000yn} by
different methods.

\subsection{Twisted vertex operators}\label{Vertices}\label{section3.4}

By considering scattering amplitudes for angled
branes,\footnote{\cite{Hashimoto:1997he, Frohlich:1999ss} performed
  related calculations in D$p$-D$q$ systems.  } it is easily seen that
the strings stretching between the branes carry the boundary condition
changing twist operators, whereas all other types of strings in the
system do not.  The vertex operators for the strings straddling the
branes represent the tachyon.  For scattering amplitude calculations,
whether the unexcited or excited twist fields are utilized depends on
the picture adopted.  However for BSFT calculations, all operators
must be expressed as worldsheet superfields, so it is most convenient
to write the relevant vertex operators as such.

Given that $\mb\twist$ has conformal dimension $\hf\nu$, an
appropriate guess for the vertex operator corresponding to the lowest
tachyonic state $\Ket0_\nu$ of figure~\ref{NSFlow} is
$$
   \mathcal{V}_{\twist^+} = T \mb\twist^+ e^{ik\cdot\mb X}
   = T\left[\twist^+_0 + i \BsCoord
     \left(\twist^+_1 - \sqrt{2\ap}\twist_0^+k\cdot\psi
     \right)\right]e^{ik\cdot X}\,.
$$ $X$ and $\psi$ are the worldsheet fields in the directions
transverse to the twisting plane.  This operator has the correct
properties to be identified with the (lowest) tachyon:
\begin{itemize}
   \item The BRST quantization condition that a vertex operator must
     have total conformal dimension $1/2$ (before ghost contributions)
     gives the state the correct mass: $\ap m_\twist^2 = \hf(\nu-1)$.
   \item When $\nu = 0$ this becomes the well-known vertex operator
     for the NS$-$ vacuum, the $\DD$ tachyon.
   \item When $\nu = 1$, since $\mb\twist \to \sD{\BsCoord}\mb Z$, the
     vertex operator becomes that of the gauge field or more
     correctly, a combination of the off-diagonal parts of the gauge
     field along one of the brane directions, $A_8$, and the
     transverse scalar, $\Phi$.  Such a linear combination of these
     fields is that state which is expected to become tachyonic from
     low energy analysis~\cite{Hashimoto:2003xz}, with other linear
     combinations being massive.  The coefficient $T$ would then be
     relabeled as $\tfrac1{\sqrt2}(A_8-i\Phi)$.
\end{itemize}
Vertex operators corresponding to other states in figures~\ref{NSFlow}
and~\ref{BoseFlow} can be constructed similarly; for instance states
in the NS$-$ spectrum at $\nu=0$ which include a $\sD{\BsCoord}\mb
Z^{(\dagger)}$ excitation (such as the second lowest state in
figure~\ref{NSFlow}) include factors of the $h = \hf - \hf\nu$
superfield
$$
   (u\sigma)^\pm - \frac i{\sqrt{2\ap}}\BsCoord(s\mu^\prime)^\pm
$$ 
which flows from $\sD{\BsCoord}\mb Z^{(\dagger)}$ to $\id$.  Again the
spectral flow is manifest in the change of statistics of the operator:
from fermionic in the NS$-$ sector to bosonic in the NS+ sector.

\section{BSFT of the brane-antibrane system}\label{section4}

Before applying our formalism to a system of branes at an angle, it is
useful to summarize the special case of brane-antibrane as described
in the context of BSFT and presented in~\cite{Kraus:2000nj,
  Takayanagi:2000rz}. We follow closely the conventions of
~\cite{Takayanagi:2000rz} and~\cite{Polchinski:1998rr}.  We first
restrict attention to D9-branes in type IIB theory.  BSFT allows one
to extract from the worldsheet sigma-model~\cite{Tseytlin:1989rr} -
deformed at the boundary with relevant ({\em i.e.} off-shell)
perturbations - the low energy effective action of spacetime fields
~\cite{Witten:1992qy, Witten:1993cr, Gerasimov:2000zp,
  Kutasov:2000qp}.  This framework developed for the bosonic
sigma-model was extended to the superstrings in~\cite{Kutasov:2000aq}
and formally justified in ~\cite{Marino:2001qc, Niarchos:2001si}.

In the NS sector the spacetime action is given in the BSFT formalism
by
 \begin{equation}
\label{DefinitionS}
   S_{\text{spacetime}} = -\int \mathcal DX\mathcal D\psi
     \mathcal D\tilde\psi\;e^{-S_\Sigma-S_{\partial\Sigma}}.
 \end{equation}
where $\Sigma$ is the worldsheet disk and $\partial\Sigma$ is its
boundary.  The worldsheet action in the bulk is typically
$$
   S_\Sigma = \frac1{4\pi}\int d^2z\;\left[
     \left(\frac2{\ap}\right) \partial X^\mu\bar\partial X_\mu 
   + \psi^\mu\bar\partial\psi_\mu 
   + \tilde\psi^\mu\partial\tilde\psi_\mu\right].
$$ The appropriate boundary insertion for different brane systems is
most easily understood through Chan-Paton factors at the string
endpoints. For instance, in the brane-antibrane system the string
endpoints couple to the superconnection~\cite{Quillen, Witten:1998cd}
and the boundary insertion is~\cite{Kraus:2000nj, Takayanagi:2000rz}
\begin{equation}
\label{BoundaryInsertion}
   e^{-S_{\partial\Sigma}} = \Tr\mb P\exp\left[
     \int d\tau d\BsCoord \mathcal M(\mb X)\right],\qquad
   \mathcal M(\mb X) = \left(\begin{array}{cc}
     iA^1_\mu(\mb X)\sD{\BsCoord}\mb X^\mu&\sqrt\ap\Tbar(\mb X)\\
     \sqrt\ap T(\mb X)&iA^2_\mu(\mb X)\sD{\BsCoord}\mb X^\mu
   \end{array}\right).
 \end{equation}
Bold quantities denote superfields, and $\mb P$ denotes supersymmetric
path ordering which is necessary to preserve worldsheet supersymmetry
and gauge invariance.  $A^{1,2}$ are the $\UU(1)$ connections, and $T$
is the tachyon charged under the relative $\UU(1)$,
$$
   D_{\mu}T = \left[\partial + iA^-\right]_{\mu}T\,,\qquad
   \hbox{where} \qquad
   A^\pm_\mu = (A^1 \pm A^2)_\mu\,.
$$ The diagonal entries of $\mathcal M$ are understood as the fields
on the brane and antibrane, and the off-diagonal entries represent the
fields associated with strings stretching between them.
Equation~(\ref{BoundaryInsertion}) is written in one dimensional
boundary superspace hence it is manifestly invariant under $\N=1$
supersymmetry.  The boundary \pagebreak[3] superfields are (see
\textref{SuAppendix}{appendix} for conventions)
$$
   \mb X^\mu = X^\mu + i\BsCoord\sqrt{2\ap}\psi^\mu\,,\qquad
   \sD{\BsCoord} = \partial_\BsCoord + \BsCoord\partial_\tau\,.
$$ 
The path ordered trace in~(\ref{BoundaryInsertion}) is most readily
performed using a complex boundary fermion superfield $\mb\Gamma =
\eta + \BsCoord F$, the quantization of which satisfies the algebra of
the $\SU(2)$ matrices~\cite{Marcus:1987cm, Marcus:1988zy}.
Hence~(\ref{BoundaryInsertion}) can be written
\begin{eqnarray}
   e^{-S_{\partial\Sigma}} &=& 
   \int \mathcal D\mb\Gamma \mathcal D\bar{\mb\Gamma}
   \exp\Biggl[\int d\tau d\BsCoord
     \Biggl(\mb\Gamma \sD{\BsCoord}\bar{\mb\Gamma}
     + \frac i2 A^+_\mu(\mb X)\sD{\BsCoord}\mb X^\mu
     + \sqrt\ap T\bar{\mb\Gamma} + \sqrt\ap\Tbar\mb\Gamma+
\nonumber\\&&
         \hphantom{   \int \mathcal D\mb\Gamma \mathcal D\bar{\mb\Gamma}
   \exp\Biggl[\int d\tau d\BsCoord
     \Biggl(}
     + iA^-_\mu(\mb X)\sD{\BsCoord}\mb X^\mu \bar{\mb\Gamma}\mb\Gamma
     \Biggr)\Biggr]\,.
 \end{eqnarray}
 When the gauge fields vanish, the boundary fermion superfields can be
 integrated to give
\begin{equation}
\label{DDbarInsertion}
   e^{-S_{\partial\Sigma}} = \exp\left[\ap\int d\tau\left(
     -T\Tbar + 2\ap (\psi\cdot\partial T)
     \frac1{\partial_\tau}(\psi\cdot\partial \Tbar) \right)\right].
 \end{equation}
The operator $1/\partial_\tau$ acting on a function $f(\tau)$ is
defined to be the convolution of $f$ with sgn$(\tau)$ over the
worldsheet boundary.  With this insertion~(\ref{DefinitionS}) remains
Gaussian for a linear tachyon profile~\cite{Kraus:2000nj,
  Takayanagi:2000rz}.  The result can be interpreted as a spacetime
action for the $\DD$ system.  The path integral can still be performed
when $A^+ \ne 0$, and given that the tachyon is charged under the
relative $\UU(1)$, there is a unique way to write the spacetime action
in a gauge covariant form~\cite{Jones:2002si}, giving
 \begin{align*}
   S_{\DD} &= -\tau_9\int d^{10}x\;e^{-2\pi\ap T\Tbar}
   \Bigg[\begin{array}{l}
       \sqrt{-\det[\G_1]}\F(\X_1+\sqrt \Y_1)\F(\X_1-\sqrt \Y_1)\\
       +\sqrt{-\det[\G_2]}\F(\X_2+\sqrt \Y_2)\F(\X_2-\sqrt \Y_2)
     \end{array}\Bigg],\\\nonumber
   (\G_{\mu\nu})_{1,2} &\equiv \left(g_{\mu\nu} 
   + 2\pi\ap F^{1,2}_{\mu\nu}\right),
   \quad\quad\quad
   \begin{array}{rl}
     \X_{1,2} &\equiv 2\pi\ap^2\G^{\{\mu\nu\}}_{1,2}
     D_\mu TD_\nu\Tbar,\\
     \Y_{1,2} &\equiv\left|2\pi\ap^2\G^{\mu\nu}_{1,2}
     D_\mu TD_\nu T\right|^2\,,
   \end{array}
 \end{align*}
where the function $\F(x)$ is the ``boundary entropy'' found in
~\cite{Kutasov:2000aq},
$$
   \F(x) = \frac{4^xx\Gamma(x)^2}{2\Gamma(2x)}
   = \frac{\sqrt{\pi}\Gamma(1+x)}{\Gamma(\hf+x)}\,.
$$ As $x \rightarrow 0$, $\F(x) \rightarrow 1 + (2 \ln2) x$, yielding
a tachyon mass $m_T^2 = -1/(4 \ln 2)$.  The factor of $2 \ln 2$ is a
well-known discrepancy between the BSFT tachyon mass and the
worldsheet CFT mass.

\section{BSFT for branes at angles}\label{AngleBSFT}\label{section5}

The most obvious way to extend the brane-antibrane BSFT formalism to a
system of branes at an arbitrary angle is to T-dualize the boundary
insertion~(\ref{BoundaryInsertion}) and turn on background values for
the scalars representing the brane positions in the extra directions.
However, this approach very quickly leads to non-Gaussian boundary
interactions, and hence does not have a free-field solution.  We find
that an approach using boundary twist superfields is more fruitful, as
was first performed in a small $(1-\nu)$, small $T\Tbar$ expansion by
~\cite{David:2000yn} and was suggested in~\cite{Martinec:2002tz}.

\pagebreak[3]

The twist superfield obtained in section~\ref{BTwist} is
$$
   \mb\twist^\pm = \twist^\pm_0 + i\BsCoord \twist^\pm_1 
   = (\sigma s)^\pm - \frac i{\sqrt{2\ap}}\BsCoord (\mu u)^\pm\,,\qquad
   h_{\twist^\pm_0} = \hf\nu\,,\qquad
   h_{\twist^\pm_1} = \hf(\nu + 1)\,,
$$ where $\nu = \phi/\pi$, with $\phi$ being the angle between the
branes.  The superfields $\mb\twist^+$ and $\mb\twist^-$ differ in
that they twist the bosons and fermions in opposite directions and
must come in pairs to give a consistent worldsheet; \ie the boundary
conditions must flip from those of one D-brane to those of the other
back and forth. The vertex superfield which interpolates between that
of the $\DD$ tachyon and that of the DD gauge field was elucidated in
section~\ref{Vertices}:
 \begin{eqnarray}
   \mathcal{V}_{\twist^+} &=& T \mb\twist^+ e^{ik\cdot\mb X}
\nonumber\\
   &=& T\left[\twist^+_0 + i \BsCoord
     \left(\twist^+_1 - \sqrt{2\ap}\twist_0^+k\cdot\psi
     \right)\right]e^{ik\cdot X}\,,
 \end{eqnarray}
where $X$ and $\psi$ are in the untwisted directions only.  The
coupling $T$ is to be interpreted as a tachyon with no momentum in the
$X^8$ direction, yet for a full twist $\nu\to1$, it is to be
interpreted as the off-diagonal parts of $\tfrac1{\sqrt2}(A_8-i\Phi)$,
where $\Phi$ is the scalar representing the brane location in the
transverse direction.  A vertex operator $\mathcal{V}_{\twist^-}$ can
be similarly constructed by flipping $\mb\twist^+$ and $\mb\twist^-$.
Exponentiation of the correlation functions of these vertex operators
leads to the sigma-model action for the lowest tachyonic state
stretching between angled branes.  Alternatively, the BSFT boundary
insertion can be constructed as the direct generalisation of
~(\ref{BoundaryInsertion}), by noting that only tachyon fields should
carry twist superfields since they are the states stretching between
branes, and that the twist superfields have the appropriate limiting
forms in terms of the bosonic superfields.  The boundary insertion
should therefore be
$$
   e^{-S_{\partial\Sigma}} = \Tr\mb P\exp\left[\int d\tau d\BsCoord 
     \mathcal M(\mb X)\right],\qquad
   \mathcal M(\mb X) = \left(\begin{array}{cc}
     0 & \sqrt\ap \mb\twist^-\Tbar(\mb X)\\
     \sqrt\ap T(\mb X) \mb\twist^+ & 0
   \end{array}\right),
$$ 
where we have set the gauge fields corresponding to strings starting
and ending on the same brane to zero for simplicity. We will comment
on the inclusion of these fields in the Discussion section.  This
modification of the brane-antibrane worldsheet action reduces to the
insertion for the $\DD$ system when $\nu=0$ where $\mb\twist^\pm$
become $\id$.  When $\nu=1$, the twist superfields become
$\sim\sD{\BsCoord}\mb Z^{(\dagger)}$, so the insertion is exactly that
for the off-diagonal fields of the brane-brane system if $T$ is
interpreted as a combination of the gauge field in one direction and
the transverse scalar.  Note that the components of these off-diagonal
gauge fields in the untwisted directions are not incorporated in this
procedure because they correspond to massive string modes for all $\nu
< 1$.  The insertion will always produce a consistent worldsheet
theory since the twist and anti-twist operators always appear
together.

The worldsheet actions we write must be regarded as formal expressions
since the treatment of the twist fields as classical fields in the
action is ill-defined; these fields have fractional statistics so can
be represented neither as complex numbers nor Grassman valued
functions.  We take these boundary actions to represent the
appropriate formal sum of correlation functions, which are well
defined for the twist fields.  In evaluating partition sums with these
boundary insertions, we will resort to correlation function methods in
order to side-step any ambiguities.  The formal action is used to
determine the equations satisfied by the correlation functions in the
interacting theories.

\pagebreak[3]

To proceed, what is usually done in the literature is to represent
path ordered traces in terms of boundary fermions.  Since we are
working with operators of fractional statistics, the boundary fermion
formalism is more intricate, so we will continue as far as possible
without recourse to it.  The supersymmetric path-ordering can be
evaluated directly from the definition as written
in~\cite{Kraus:2000nj}.  We set the gauge fields to zero for
simplicity and obtain
\begin{eqnarray}
\nonumber
   && \Tr\mb P\exp\int d\hat\tau \mathcal M(\hat\tau)
   = \Tr\sum_{n=0}^\infty\int d\hat\tau_1\ldots d\hat\tau_n
   \Theta(\hat\tau_{12})\Theta(\hat\tau_{23})
   \ldots\Theta(\hat\tau_{n-1,n})
   \mathcal M(\hat\tau_1)\ldots\mathcal M(\hat\tau_n)
\\
\nonumber
   &&\qquad =\Tr\sum_{n=0}^\infty\int d\tau_1\ldots d\tau_n\left[
   \begin{array}{l}
     \Theta(\tau_{12})\Theta(\tau_{23})\ldots\Theta(\tau_{n-1,n})
     \times\\
     \left[\mathcal M_1-\mathcal M_0^2\right](\tau_1)\ldots
     \left[\mathcal M_1-\mathcal M_0^2\right](\tau_n)
   \end{array}\right]\\\nonumber
   &&\qquad= \Tr P\exp\int d\tau\left[
     \mathcal M_1(\tau) - \mathcal M_0^2(\tau)\right]\\
   &&\qquad= \Tr P\exp\left[\ap\int d\tau
     \left(\begin{array}{cc}
       -\twist_0^-\Tbar T\twist_0^+&
       i\sqrt{2\ap}\twist_0^- \partial_i \Tbar\psi^i 
       + i\twist_1^-\Tbar\\
       i\sqrt{2\ap}\partial_j T\psi^j \twist_0^+ + i T\twist_1^+&
       - T\twist_0^+\twist_0^-\Tbar
     \end{array}\right)\right],\label{SuPathOrder}
 \end{eqnarray}
in which $\hat\tau_{12} = \tau_1-\tau_2+\BsCoord_1\BsCoord_2$, $P$ in
the result is standard path ordering after integration over
superspace, and $\mathcal M_{0,1}$ are the parts of the matrix
$\mathcal M$ which are proportional to zero and one power of
$\BsCoord$ respectively.  The path ordered trace is most simple to
evaluate using boundary fermions, but in the present case cocycles are
needed because the twist fields have fractional statistics.
Proceeding by assuming such cocycles can be defined, we write the
simplified boundary insertion after taking the path ordered trace by
modified boundary fermions as
 \begin{equation}
   e^{-S_{\partial\Sigma}} 
   = \exp\int \ap d\tau\Biggl[-\twist^-_0\Tbar T\twist^+_0 
     + \left(\sqrt{2\ap}\twist^-_0\partial \Tbar\cdot\psi
     + \twist^-_1 \Tbar \right) \frac1{\partial_\tau}
     \left(\sqrt{2\ap}\partial T\cdot\psi\twist^+_0
     + T \twist^+_1 \right)
     \Biggr]\,.\label{TwistedBAction}
 \end{equation}
The $T$ constant version of this worldsheet action was derived in
~\cite{David:2000yn} using boundary fermions, but that derivation
assumed $\twist_0$ and $\twist_1$ are always bosonic and fermionic
respectively.  Alternatively we can construct this formal insertion as
the simplest expression which satisfies the necessary properties that
\begin{itemize}
   \item it exhibits explicit $\N = 1$ 1D SUSY,
   \item boundary condition changing operators always appear in
     $\twist^-\twist^+$ pairs, for a consistent worldsheet.  Although
     individual twist fields are fractionally moded, the $\twist^-
     \twist^+$ pairs behave as scalars.
   \item in the limit $\nu \to 0$, because we have $\twist_0\to\id$
     and $\twist_1\to0$, the $\DD$ boundary insertion for the tachyon
     is exactly reproduced,
   \item for $\nu\to1$, the boundary action becomes that for gauge
     fields and scalars on the brane-brane system.  For instance, when
     $\partial T$ is set to zero and $\nu = 1$ we have
   $$
     S_{\partial \Sigma} \to \int d\tau \left(-\frac i2 T\Tbar\right) 
     \left[X^8(X^9)^\prime - 2\ap \psi^8\psi^9 \right].
   $$ 
If we relabel the coupling $T\Tbar \to i(\bar\Phi A_8 - \bar A_8
\Phi)$ (which is real, with $A_8$ and $\Phi$ the complex off-diagonal
parts of the $\UU(2)$ gauge field and brane scalar), this is exactly
the correct boundary insertion for the brane brane system after
setting all field derivatives and all other fields in the system to
zero; \ie this is just the insertion for that part of $D\Phi$ which we
have not set to zero, and in this case the path ordered trace is
trivial.  All other fields were set to zero since they are massive for
$\nu < 1$, or because we neglected field derivatives.
\end{itemize}
As such, we treat~(\ref{TwistedBAction}) as a well-motivated
definition of the boundary insertion for the tachyonic state for
branes at angles, even though we are unable to give a thorough
derivation of it.

\subsection{Tachyon potential}\label{section5.1}

The tachyon potential is obtained by inserting $T(x) = T$ (constant)
on the boundary.  The non-trivial $\nu$ dependence is completely given
by the twist fields.  The potential is
\begin{equation}\label{NuPotential}
   V_\nu(T\Tbar) = \int \mathcal DX\mathcal D\psi\; e^{-S_\Sigma}
   \exp\int d\tau (-\ap T\Tbar) \left[\twist_0^-\twist_0^+ 
     - \twist_1^-\frac 1{\partial_\tau}\twist_1^+ \right].
 \end{equation}
Because of the ambiguity of treating the twist fields as classical
worldsheet fields, the path integral cannot be performed directly.
Since all twist field correlators are well defined however, this
potential can be evaluated by identifying the equations they satisfy
in the presence of the boundary insertion.  Writing $T\Tbar = y$, we
see that
 \begin{eqnarray}
   \partial_y \ln V_\nu(y) &=& -\ap\int d\tau\left\{
     \frac{\Expect{\twist_0^-(\tau)\twist_0^+(\tau)}_y}{\Expect{1}_y}
     - \frac{\Expect{\twist_1^-(\tau)
	 \frac1{\partial_\tau}\twist_1^+(\tau)}_y}{\Expect{1}_y}
     \right\}
\nonumber\\
   &=& -\frac\ap2 \int d\tau\left\{ G^-(0;y) + G^+(0;y)
     +\frac1{\partial_\epsilon}\left[ H^-(0;y) + H^+(0;y) 
     \right]\right\}.
\label{VEqn}
 \end{eqnarray}
The $y$-dependent correlators are defined as
$$
   \Expect{\mathcal O}_y \equiv \Expect{\mathcal O 
     \exp\int d\tau (-\ap y) \left[\twist_0^-\twist_0^+ 
     - \twist_1^-\frac 1{\partial_\tau}\twist_1^+ \right]},
$$ so $V_\nu(y) = \Expect{1}_y$, and we have defined the point-split
correlators of twist fields
 \begin{equation}
   G^\mp(\epsilon;y) \equiv 
   \frac{\Expect{\twist_0^\mp(\epsilon)\twist_0^\pm(0)}_y}{\Expect{1}_y}\,,
\qquad
   H^\mp(\epsilon;y) \equiv  
   \frac{\Expect{\twist_1^\mp(\epsilon)\twist_1^\pm(0)}_y}{\Expect{1}_y}\,,
   \label{GreensDefs}
 \end{equation}
with $\epsilon\rightarrow 0^+$.  As in~\cite{Kutasov:2000aq,
  Takayanagi:2000rz} we shall use a point-splitting method to evaluate
the action.\footnote{One can see that equation~(\ref{VEqn}) involves
  both orderings as the limit $\epsilon\rightarrow 0^+$ is taken by
  looking at the original regularization prescription adopted
  in~\cite{Kutasov:2000aq, Takayanagi:2000rz} and noting that our
  fields are complexified; \ie $\Expect{X^8(\epsilon) X^8(0) +
    X^9(\epsilon) X^9(0)} = \Expect{Z(\epsilon)\Zb(0) +
    \Zb(\epsilon)Z(0)}$.}  By differentiating~(\ref{GreensDefs}), the
ordered Green's functions are easily seen to satisfy the differential
equations
 \begin{eqnarray}
\nonumber
   \partial_y G^\mp(\epsilon;y) &=& 
   -\ap\int d\tau\; G^\pm(\tau-\epsilon;y) G^\mp(\tau;y)\,,
\\
   \partial_y H^\mp(\epsilon;y) &=& 
   -\ap\frac1{\partial_\epsilon}\int d\tau 
   H^\pm(\tau-\epsilon;y)H^\mp(\tau;y)\,.\label{GHEqns}
 \end{eqnarray}
A problem with these equations is that they are not valid for the
$\DD$ case, $\nu = 0$, for which $\Expect{1}_y \sim \exp(-2\pi\ap y)$
and $G^\pm(y) = 1$.  This will cause a discontinuity in the
$\nu$-dependent solution.  However, the cause of the discontinuity is
well understood - the operators $\twist^\mp\twist^\pm$ merge in the
limit $\nu\to0$, and so~(\ref{GHEqns}) includes too many Wick
contractions only for that case.  Also we find $\lim_{\nu\to0}
V_\nu(y)$ is not very different from the $\DD$ potential, and leads to
the same physical behaviour.

These equations can be easily solved if one expands the Green's
functions of the interacting theory $G^\pm$ and $H^\pm$ in frequency
modes in a manner consistent with the ordering prescription adopted
earlier.  In particular, $G^-$ and $H^-$ are expanded in positive
modes, while $G^+$ and $H^+$ are negatively moded.  The \emph{ansatz}
for the correlators is
$$
   G^\mp(\epsilon;y) = \sum_{n=0}^\infty
   \mathcal G^\mp_n(y)e^{\pm i\epsilon[n+\hf\nu]}\,,\qquad
   H^\mp(\epsilon;y) = \sum_{n=0}^\infty
   \mathcal H^\mp_n(y)e^{\pm i\epsilon[n+\hf(\nu+1)]}\,.
$$ 
The equations which the mode coefficients satisfy, and their boundary
conditions from the free theory, $y=0$, correlators of
section~\ref{TFCorrelators} are
\begin{equation}
\begin{array}[b]{rclcrcl}
   \displaystyle\frac{d}{dy}\mathcal G^-_n(y) &=& \displaystyle\frac{d}{dy}\mathcal G^+_n (y)
   = -2\pi\ap \mathcal G^+_n(y) \mathcal G^-_n(y)\,,&\qquad&
   \displaystyle\mathcal G^\mp_n(0) &=& \displaystyle\frac{e^{\mp i\frac\pi2\nu}}{n\B(n,\nu)}\,,
\\[7pt]
   \displaystyle\frac{d}{dy}\mathcal H^-_n(y) &=&
   \displaystyle-\frac{d}{dy}\mathcal H^+_n(y) = \frac{2\pi\ap
     i}{n+\hf(\nu+1)} \mathcal H^+_n(y) \mathcal H^-_n(y)\,, &\qquad&
   \displaystyle\mathcal H^\mp_n(0) &=& \displaystyle\frac{\mp ie^{\mp
       i\frac\pi2\nu}}{\B(n+1,\nu)}\,,
 \end{array}
\end{equation}
which have solutions
 \begin{eqnarray}
   \mathcal G^\mp_n(y) &=& \frac{\sin(\tfrac{\pi\nu}2)}{n\B(n,\nu)}
   \left[\cot\left(\frac{\sin(\tfrac{\pi\nu}2)}{n\B(n,\nu)}2\pi\ap y 
     + \tfrac{\pi\nu}2\right) \mp i\right],
\nonumber\\
   \mathcal H^\mp_n(y) &=& -\frac{\sin(\tfrac{\pi\nu}2)}{\B(n+1,\nu)}
   \left[1 \pm i \cot\left(\frac{\sin(\tfrac{\pi\nu}2)}
     {[n+\hf(\nu+1)]\B(n+1,\nu)}2\pi\ap y  +
   \tfrac{\pi\nu}2\right)\right].
 \end{eqnarray}
Finally, we can reconstruct the potential using~(\ref{VEqn}) and these
solutions.
 \begin{equation}\label{nuPotential}
   V_\nu(y) = N_\nu \prod_{n=0}^\infty \frac
   {\displaystyle\sin\left(\frac{\sin(\tfrac{\pi\nu}2) 2\pi\ap y}
     {[n+\hf(\nu+1)]\B(n+1,\nu)} + \frac{\pi\nu}2\right)}
   {\displaystyle\sin\left(\frac{\sin(\tfrac{\pi\nu}2)}{n\B(n,\nu)}2\pi\ap y 
     + \frac{\pi\nu}2\right)}.
 \end{equation}
 $N_\nu$ is the unknown normalization which we shall discuss in
 section~\ref{LEEASec}.  Note that $V_\nu(0) = N_\nu
 [\sin(\tfrac{\pi\nu}2)]^{-\hf}$, where $\zeta$-function regularization
 must be used to evaluate the product at $y=0$. ~(\ref{nuPotential})
 leads to the $\DD$ and DD potentials,
$$
   V_0(T) = \frac{V_0(0)}{1+2\pi\ap T\Tbar}\,,\qquad
   V_1(T) = \frac{V_1(0)}{\sqrt{\cos(2\pi\ap T\Tbar)}}\,.
$$ 
In the $\nu = 0$ limit only the $n=0$ term from the product has $y$
dependence.  The reason for the discrepancy from the expected tachyon
potential, $\exp(-2\pi\ap T\Tbar)$, was explained previously; these
two functions have similar behaviour so we do not consider this
discontinuity a serious problem.  Na\"{\i}vely the $\nu=1$ result is
constant since the numerator and denominator at each $n$ are
identical, but we must take into account that the numerator comes from
worldsheet modes of frequencies $\ints + \tfrac{\nu + 1}2$, while the
denominator from modes of frequencies $\ints + \nu/2$; using
$\zeta$-function regularization as in the derivation of the DBI
~\cite{Tseytlin:1999dj},
$$
   \frac{V_1(T)}{V_1(0)} = \frac{
     \prod_{m = 1}^\infty \cos(2\pi\ap T\Tbar)}
   {\prod_{r = \tfrac12}^\infty \cos(2\pi\ap T\Tbar)}
   = \left[\cos(2\pi\ap T\Tbar)\right]^{
   \zeta(0,1) - \zeta(0,1/2)}
   = \left[\cos(2\pi\ap T\Tbar)\right]^{-1/2}\,.
$$

\FIGURE[t]{\centerline{\epsfig{file = nuPotential.eps, width = 7cm,clip=}}%
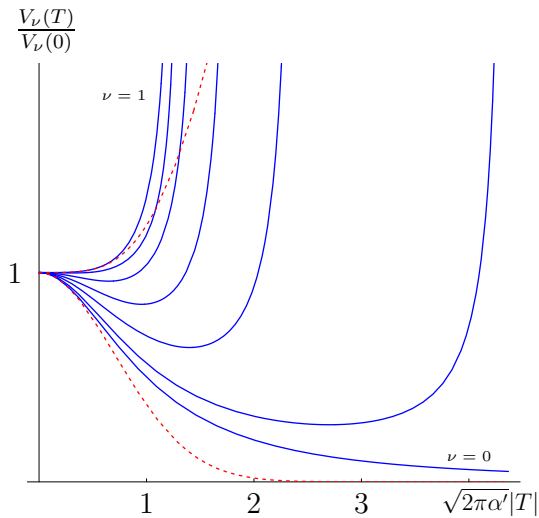
\caption{Solid curves show $V_\nu(T)$ for $\nu = 0, 0.1, 0.3, 0.5,
  0.7, 0.9$ and $1$.  The dotted curves show the gaussian $\DD$
  tachyon potential and the DBI, the expected results for $\nu = 0$
  and $\nu=1$.\label{VnuPlot}}}
 
That this differs from the known result, $\sim\sqrt{1 +
  \tfrac14(2\pi\ap T\Tbar)^2}$ is not yet understood, but again the
two potentials shall give similar physical behaviour.  The
regularization by $\zeta$-function methods means that we must take
infinitely many terms into account for the $\nu = 1$ case; as $\nu: 0
\to 1$ more and more terms in the product~(\ref{nuPotential}) become
relevant.  This can also be seen by considering that the frequencies
of the sine functions for higher terms are less than those for lower
$n$, so the higher $n$ terms tend to be constant over the range of
physical interest.  As $\nu$ approaches 1, the frequencies of all
terms approach one value, so the higher terms vary more and more.
Note also that a tower of (infinite number of) massive states becomes
massless at $\nu=1$. They are not included in the above calculation.

Figure~\ref{VnuPlot} shows that for intermediate values of $\nu$, the
potential has appropriate properties for branes at angles.  Firstly,
we see that the minimum of $V_\nu$ is at $T\to\infty$ for $\nu=0$, and
smoothly moves to finite values for intermediate $\nu$, to approach $T
= 0$ at $\nu=1$.  The potential also approaches a singularity at
$2\pi\ap T\Tbar = \pi(1-\tfrac\nu2)/\sin\tfrac{\pi\nu}2$.  This is not
problematic given that the tachyonic state is localized about the
intersection and the tachyon condensation must end at some finite
expectation value, as will be discussed in section~\ref{LEEASec}.
That the potential is an infinite product of poles and zeros can be
disconcerting, but it is always well-behaved in the region of physical
interest because the frequencies of the higher modes decrease - so the
potential always reaches the singularity from the $n=0$ term sooner
than it reaches any other zero or pole.

The mass term can also be obtained from~(\ref{nuPotential}):
 \begin{eqnarray}\nonumber
   \tilde m_\twist^2(\nu) \equiv \partial_yV_\nu(y)|_{y = 0}
   &=& 2\pi\ap\cos\frac{\pi\nu}2
   \sum_{n=0}^\infty \frac{\Gamma(n+\nu)}{\Gamma(n+1)\Gamma(\nu)}
   \left[\frac{n+\nu}{n+\hf(\nu+1)}-1\right]
\\
   &=& -\pi\ap\cos\frac{\pi\nu}2 (1-\nu)
   \frac{\Gamma(1-\nu)\Gamma(\tfrac{1+\nu}2)}{\Gamma(\tfrac{3-\nu}2)}\,.
   \label{BareMass}
 \end{eqnarray}
This is always negative and vanishes at $\nu = 1$, as can be seen in
figure~\ref{VnuPlot}.  It is not the physical mass of the twisted
tachyon state however, since the kinetic term is not canonically
normalized; in the section to follow, we calculate the normalization
of the kinetic term to extract the physical tachyon mass.

\subsection{Tachyon kinetic terms}\label{section5.2}
 
Unlike in the brane-antibrane case, the tachyon potential is the only
quantity which can be evaluated exactly using BSFT.  This is because
in order to angle the one brane with respect to the other, additional
dynamical worldsheet fields need to be included in the
action~(\ref{TwistedBAction}), and this means that a linear tachyon
profile corresponds to a non-Gaussian path integration.  Since the
coupling of the linear tachyon profile corresponds in the low energy
action to the tachyon kinetic term, the complete kinetic term cannot
be obtained for branes at angles.  We can, however, calculate the
tachyon mass by performing an expansion of~(\ref{TwistedBAction}) in
$\partial_i T$; the coefficient of the term of $(\partial_i T)^2$ then
gives the normalization of the kinetic term in the low energy
effective action, which shall allow us to determine the mass of the
lowest tachyon.

With $T$ linear in one of the un-angled directions along the brane,
$T=uX$, the coefficient of $-2\pi\ap u^2$ in~(\ref{TwistedBAction}) is
 \begin{eqnarray}
   && \Expect{\twist_0^-X(\tau_1)X\twist_0^+(\tau_2)}
   -\frac1{\partial_{\tau_2}}
   \Expect{\left[\sqrt{2\ap}\twist_0^-\psi + \twist_1^-X\right]\!(\tau_1)
   \left[\sqrt{2\ap}\psi\twist_0^+ + X\twist_1^+\right]\!(\tau_2)}=
\nonumber\\
   \qquad &&= \underbrace{
     \Expect{\twist_0^-\twist_0^+}\!(\epsilon)\Expect{XX}\!(\epsilon)
   }_{2\ap \mathcal C_0(\epsilon)}
   + \frac1{\partial_\epsilon}\Biggl[2\ap\underbrace{
     \Expect{\twist_0^-\twist_0^+}\!(\epsilon)\Expect{\psi\psi}\!(\epsilon)
   }_{\mathcal C_1(\epsilon)}
   + \underbrace{
     \Expect{\twist_1^-\twist_1^+}\!(\epsilon)\Expect{XX}\!(\epsilon)
   }_{2\ap \mathcal C_2(\epsilon)}\Biggr]\,,
\nonumber
 \end{eqnarray}
where $\epsilon = \tau_1-\tau_2\to0$.  The $\nu$-dependent coefficient
of the kinetic term for the tachyon will therefore be given by
 \begin{equation}
\label{KineticFinite}
   \mathcal K_\nu \equiv -4\pi\ap^2\left\{
   \mathcal C_0(0) + \frac1{\partial_\epsilon}\left[
     \mathcal C_1 + \mathcal C_2 \right](0)\right\}.
 \end{equation}
To calculate this quantity we expand the correlators in modes, and set
$\epsilon = 0$.  The individual sums obtained diverge but the
combination is finite.  Formally, in order to obtain the mode
expansions, we add a small negative imaginary component to $\epsilon$.
As in the previous section, we average over the two \pagebreak[3] physically
distinct point-splitting schemes, giving both positive and negative
frequency modes.
\begin{eqnarray}
   \mathcal C_0(\epsilon) &=& \frac{
     -\ln\left(2\sin\frac\epsilon2\right)}{
     \left[2\sin\frac\epsilon2\right]^\nu}
   = \hf \sum_{\substack{m=1\\n=0}}^\infty
   \frac{
      e^{-i\tfrac\pi2\nu} e^{i\epsilon[n+m+\hf\nu]}
     + e^{i\tfrac\pi2\nu} e^{-i\epsilon[n+m+\hf\nu]}
     }{mn\B(n,\nu)}\,,
\nonumber\\
   \mathcal C_1(\epsilon) &=& \frac1{
     \left[2\sin\frac\epsilon2\right]^{\nu+1}}
    = \hf\sum_{\substack{m=0\\n=0}}^\infty
   \frac{
     e^{-i\tfrac\pi2(\nu+1)}e^{i\epsilon[n+m+\hf(\nu+1)]}
     + e^{i\tfrac\pi2(\nu+1)}e^{-i\epsilon[n+m+\hf(\nu+1)]}
     }{n\B(n,\nu)}\,,
\nonumber\\
   \mathcal C_2(\epsilon) &=& \frac{
     -\nu\ln\left(2\sin\frac\epsilon2\right)}{
     \left[2\sin\frac\epsilon2\right]^{\nu+1}}
   = \frac\nu2 \sum_{\substack{m=1\\n=0}}^\infty
   \frac{
     e^{-i\tfrac\pi2(\nu+1)}e^{i\epsilon[n+m+\hf(\nu+1)]}
     + e^{i\tfrac\pi2(\nu+1)}e^{-i\epsilon[n+m+\hf(\nu+1)]}
     }{n\B(n,\nu+1)}\,.
\nonumber
 \end{eqnarray}
The required combinations are the individually divergent sums
\begin{eqnarray}
   \mathcal C_0(0)
   &=& \cos\left(\frac{\pi\nu}{2}\right)\sum_{\substack{m=1\\n=0}}^\infty
   \frac{\Gamma(n+\nu)}{m\Gamma(\nu)\Gamma(n+1)}\,,
\nonumber\\
   \frac1{\partial_\epsilon}\mathcal C_1(0)
   &=& -\cos\left(\frac{\pi\nu}{2}\right)\sum_{\substack{m=0\\n=0}}^\infty
     \frac{\Gamma(n+\nu)}{[n+m+\hf(\nu+1)]\Gamma(\nu)\Gamma(n+1)}\,,
\nonumber\\
   \frac1{\partial_\epsilon}\mathcal C_2(0)
   &=& -\cos\left(\frac{\pi\nu}{2}\right)\sum_{\substack{m=1\\n=0}}^\infty
     \frac{\Gamma(n+\nu+1)}{m[n+m+\hf(\nu+1)]\Gamma(\nu)\Gamma(n+1)}\,.
\nonumber
 \end{eqnarray}
The normalization of the kinetic term~(\ref{KineticFinite}) is
obtained by performing the subtractions.  The then convergent sums can
be re-expressed as convergent integrals:
 \begin{eqnarray}
   \mathcal K_\nu &=& -4\pi\ap^2\cos\left(\frac{\pi\nu}{2}\right)
   \sum_{n=0}^\infty 
   \frac{\Gamma(n+\nu)}{\Gamma(\nu)\Gamma(n+1)}
   \left\{ -\frac1{n+\hf(\nu+1)} 
   + \sum_{m=1}^\infty \frac{\hf(1-\nu)}{m[n+m+\hf(\nu+1)]} 
   \right\}
\nonumber\\
   &=& -4\pi\ap^2\cos\left(\frac{\pi\nu}{2}\right)
   \int_0^1dx\;(1-x)^{-\nu}x^{\hf(\nu-1)}
   \left\{ -1 - \hf(1-\nu)\ln(1-x) \right\}.
\nonumber
 \end{eqnarray}
Finally, the integrals are expressible in terms of gamma and Polygamma
functions $\Psi(y)=\tfrac d{dy}\ln\Gamma(y)$,
 \begin{eqnarray}
\nonumber
   \mathcal K_\nu &=& -4\pi\ap^2\cos\left(\tfrac{\pi\nu}2\right)
   \frac{\Gamma(1-\nu)\Gamma(\tfrac{1+\nu}2)}{\Gamma(\tfrac{3-\nu}2)}
   \left\{-1 + \hf(1-\nu)\left[\Psi(\tfrac{3-\nu}2)
   -\Psi(1-\nu)\right]\right\}
\\
   &=& 2\pi\ap^2\frac{\sqrt\pi\Gamma(1-\frac\nu2)}
	{2^\nu\Gamma(\frac{3-\nu}2)}\left\{1 + (1-\nu)
	  \left[\Psi(2-\nu) 
	  -\Psi(\tfrac{3-\nu}2)\right]\right\}.\label{KResult}
\end{eqnarray}

\FIGURE[t]{\centerline{\epsfig{file=TMass.eps,width=.5\textwidth,clip=}}
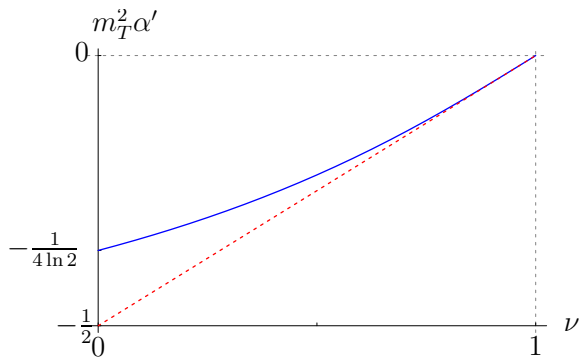
\caption{\label{TMass} The $\nu$-dependent tachyon mass calculated via
  BSFT (solid curve) and that mass calculated from worldsheet CFT
  (dotted line).  }}

In the last line $\Gamma$ and $\Psi$-function identities have been
used to write the result in a form which shows that $\mathcal K_\nu$
is finite and non-zero for $\nu\in [0,1]$.  Using this normalization
for the kinetic term and the non-nor\-ma\-li\-zed
mass~(\ref{BareMass}) obtained from the tachyon potential, the
$\nu$-dependent mass is shown in figure~\ref{TMass}. The tachyon mass
at $\nu=0$ agrees with that in the $\DD$ system in BSFT.

\subsection{Low energy effective action}\label{LEEASec}\label{section5.3}

The low energy effective action for the tachyon between two D8-branes
at angles, for which the tachyon is localized about the intersection
and has only kinetic terms in the un-angled directions is given by
 \begin{equation}
   S(T) = \int d^8x\; V_\nu(T)\sqrt{-\text{P}[g]}
   \left( 1 + \mathcal K_\nu\;\partial_aT \partial^a\Tbar \right)
   + \cdots\,,\label{LEEA}
 \end{equation}
where `$a$' runs over spacetime directions transverse to $Z$ and
$\Zb$.  The normalization of the action, which measures brane tension
and which we will now fix, is factored into $V_\nu(T)$.

It is clear from energetic considerations that the strings stretched
between two angled branes are localized about the intersection point.
It is also obvious that at the special angles of $\nu = 0$ and 1 those
strings are no longer localized.  The region of string localization
must cover the entire system for $\nu = 0$, reduces to a slight region
about the intersection, and increase again beyond $\nu = 1/2$ to once
more cover the entire system at $\nu = 1$. The localization must also
be symmetric about $\nu = 1/2$ because this argument does not depend
on brane charge.  Localization is also strongly suggested by the
spectrum, as in figure~\ref{BoseFlow} and the discussion following it.

We can estimate the size of the region of localization as a function
of $\nu$ from a simple geometric argument (see figure~\ref{Rec3}).
For a brane configuration with $\nu<1/2$, the geometric volume of
interest along the brane will be $v_1 \sim
[\sin(\tfrac{\pi\nu}2)]^{-1}$.  For $\nu>1/2$, this length is $v_2
\sim [\cos(\tfrac{\pi\nu}2)]^{-1}$.  To produce an estimate which
possesses the required symmetry about $\nu = 1/2$, the localization
volume must be an average of these two lengths, and we find that the
geometric average $\sqrt{v_1v_2} \sim [\cos(\tfrac{\pi\nu}2)
  \sin(\tfrac{\pi\nu}2)] ^{-1/2}$ matches well the result that was
calculated from a low energy analysis in~\cite{Hashimoto:2003xz}.

We expect that $V_\nu(0)$ is proportional to the volume along the
brane where the tachyon is localized.  This is sensible because we see
from figure~\ref{VnuPlot} that for intermediate values of $\nu$, the
relative potential has minimum at $0\le V_\nu(T)/V_\nu(0) \le 1$;
physically, because only a small part of the branes play a role in
recombination, if the potential were proportional to the infinite
brane volume (as in the DD and $\DD$ cases), there would be no change
in the relative energy during recombination (except when $\nu = 0$)
and the minimum of the potential would not change smoothly as it
does. The potential must therefore be proportional only to a finite
part of the brane volume, and it is sensible to assume that that part
is the volume of tachyon localization.  By matching $V_\nu(0)$ with
the energy contained in the region of localization, the normalization
of the potential, $N_\nu$ of~(\ref{nuPotential}), is determined to be
 \begin{equation}
   N_\nu = 2\tau_8 \sqrt{\frac{2\pi\ap}{\cos(\tfrac{\pi\nu}2)}}\,,\qquad
   \Longrightarrow V_\nu(0) = 2\tau_8 
   \sqrt{\frac{2\pi\ap}{\cos(\tfrac{\pi\nu}2)\sin(\tfrac{\pi\nu}2)}}\,.
 \end{equation}
Furthermore, this argument implies that the $X^8$ direction has
already been integrated out in the derivation of the potential, so the
low energy effective action is an action for dynamics in the 8
remaining spacetime dimensions of the 9-dimensional brane world
volume.

In the boundary insertion evaluated above, we consider a tachyon field
with dependence on the transverse coordinates $\mb X$. We have in mind
the lowest lying tachyonic state of the spectrum $\Ket0_\nu$.  The
dynamics also involves operators of higher conformal weights, some of
which will be relevant operators (depending on how close we are to
$\nu=0$ as argued at the end of section~\ref{SecSpec}); others will
correspond to massive modes or irrelevant perturbations. In this work,
we truncate to the cross-section (in worldsheet coupling space)
corresponding to a single complex tachyon field.  Furthermore, it is
easy to see from energetics that all string modes stretched between
the branes would be localized near the brane intersection point.
Therefore, from our BSFT formalism, the brane worldvolume accessible
to these strings is transverse to the angling plane $Z$-$\Zb$.  It is
however sometimes useful --- in particular when $\nu\sim 0$ or
$\nu\sim 1$ --- to think of a more general tachyon field with a
profile with respect to say $Z+\Zb$, with this combination of $Z$ and
$\Zb$ being considered as part of the worldvolume of the brane system.
To describe the dynamics in these variables, we may write a new field
$T(X,Z,\Zb)=\sum_{n=0} T_n(X) f_n(Z+\Zb)$ with the expansion modes
corresponding to the states $(\alpha^\dagger_{-\nu})^n \Ket0_\nu$,
with conformal dimension $((2n+1)\nu-1)/2$.  Unlike standard toroidal
Kaluza Klein compactification, the complete set given by $f_n(Z+\Zb)$
is necessarily a set of functions localized about $Z=0$. For example,
it was shown in~\cite{Hashimoto:2003xz} through a low energy treatment
at small angles, that the lowest mode $f_0(Z+\Zb)$ is a Gaussian
centered at $Z=0$.  The action for the spacetime field $T(X,Z,\Zb)$
would also have an additional integral along the worldvolume direction
$Z+\Zb$. However, in our formalism, we obtain by construction an
action with this worldvolume direction already integrated out.

Truncating to the lowest tachyon field, we are choosing to perturb
with the simpler profile $T(X,Z,\Zb)=T_0(X) f_0(Z+\Zb)$ to capture
part of the dynamics.  Normalizing the kinetic term of the new tachyon
field canonically, we can argue that the field $T_0(X)$ is
proportional to our $T(X)$ after integrating out the new worldvolume
direction; this naturally can change the form of the potential between
the two pictures. What matters however is the effective potential
written in terms of $T(X)$.  $T(X)$ may then be viewed as an average
measure of the separation between the branes - averaged over the
profile in the $Z+\Zb$ direction.  More conveniently and with the
appropriate rescaling, we may think of it as $T(X,Z=0)$; \ie the
separation between the branes at their point of nearest proximity.
This leads us to believe that our perturbation tracks the dynamics of
the more general tachyon field to leading order in resolution in $Z$
and $\Zb$.  To resolve more of the dynamics in the twisted directions,
we would then need to insert in the partition function vertex
operators for the higher modes of the string. Note that a similar
tower of massive states becomes massless at $\nu=1$.

\section{Brane recombination}\label{Recombination}\label{section6}

In this section, we present a preliminary analysis of the physics
encoded in the tachyon potential~(\ref{nuPotential}). We leave a more
detailed study of the recombination mechanism to future work.

As argued earlier, we expect that the strings stretched between two
angled branes are necessarily localized at the intersection point for
$\nu\neq 0,1$.  In the formalism we have adopted, we have perturbed
the boundary of the worldsheet with the lowest lying tachyon state -
corresponding to, in the language of~\cite{Hashimoto:2003xz}, a
Gaussian tachyon profile along the branes in the angling plane with
spread size given by $\sim [\cos(\tfrac{\pi\nu}2)
  \sin(\tfrac{\pi\nu}2) ]^{-1/2}$.  As argued in the previous section,
our boundary insertion may be thought of as tracking the magnitude of
this Gaussian.  In this context, the constant tachyon value is
physically the separation distance between the branes at their point
of closest proximity as shown in figure~\ref{Rec3}.  This is clear
near $\nu = 1$ where the tachyon is explicitly some linear combination
of the off-diagonal scalars representing brane separation and the
off-diagonal gauge field; we assume this correspondence holds for all
angles.

 \DOUBLEFIGURE[t]{TMin.eps}{VMin.eps}{
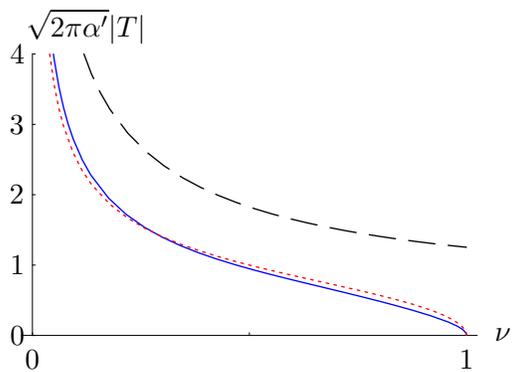
   The value of $|T|$ for which $V_\nu(T)$ is minimal (solid), the
   approximation to it, $\sqrt{2\pi\ap} \langle|T|\rangle =
   \sqrt{\cot(\tfrac{\pi\nu}2)}$ (dotted), and the critical value of
   $|T|$ for which the potential reaches a pole (dashed).
   \label{TMinPlot}}{
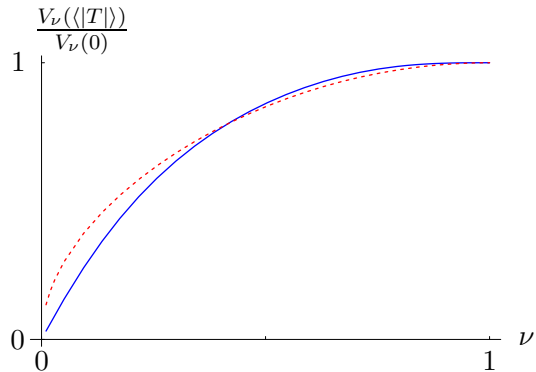
   The value of the relative potential at its minimum (solid) and the
   approximation to it $\sim \sqrt{\sin(\tfrac{\pi\nu}2)}$ (dotted), as
   functions of $\nu$.
   \label{VMinPlot}
 }
 
An interesting novel feature of our potential is that, for generic
$\nu$, the potential has a minimum for a finite expectation value of
the tachyon.  This suggests that the recombined branes separate up to
a fixed extent as the tachyon condenses.  The resultant configuration
will then be two bent branes separated by more than a string length
and so not interacting by open string modes.  The configuration will
clearly evolve from that stage; classically, and to this level of
approximation, there will be oscillation about this configuration of
bent branes (if the system is compactified), yet at this new vacuum,
the original tachyon field has become stable.
In~\cite{Hashimoto:2003xz}, it was argued that the small angle
analysis at low energies is valid until the tachyon reaches an
expectation value of $\Expect{|T|} \sim \sqrt{\cot(\tfrac{\pi\nu}2)}$;
beyond this point, the shape of the recombined branes begins to kink
and the leading low energy approximation breaks down. At this
$\Expect{|T|}$, energy loss can be calculated by integrating the shape
of the recombined brane.  This gives as the relative energy remaining
$\sim \sqrt {\sin(\tfrac{\pi\nu}2)}$.  Note that this is the square
root of the same ratio when the final state is two straight parallel
branes, giving credence to the interpretation of the final state as
bent branes.

We can calculate these quantities numerically from our
potential~(\ref{nuPotential}) assuming that $V_\nu(T)$ is proportional
only to that part of the brane which plays some role in the
recombination; \ie $V_\nu(0)$ is proportional to the volume along the
brane in which the tachyon is localized.  The results are shown in
figures~\ref{TMinPlot} and \ref{VMinPlot}.  We find that the numerical
values of the stable tachyon expectation value and the potential at
its minimum are to an excellent approximation given by
\begin{equation}
   \sqrt{2\pi\ap}\Expect{|T|} \sim \sqrt{\cot\left(\frac{\pi\nu}2\right)},\qquad
   \frac{V_\nu(\Expect{|T|})}{V_\nu(0)} 
   \sim \sqrt{\sin\left(\frac{\pi\nu}2\right)},
\end{equation}
in good agreement with the $\nu$-dependence of the low energy
estimates of these quantities at the regime where the low energy
approximation breaks down.

To summarize the interpretation of these results, the fact that
$V_\nu(T)$ has a minimum at finite $\Expect{|T|}$ suggests that the
tachyonic mode is localized about the intersection point, as expected
from other physical considerations, and that the potential is
proportional to the localization volume.  This tachyonic mode drives
the dynamics of the brane recombination to a certain separation,
beyond which the system evolves as two bent branes with only massive
open string (and closed strings) interactions between them.

\section{Discussion}\label{section7}

In this work, we have developed the formalism of twist fields, which
are frequently employed in the study of orbifolds, to the BSFT of
angled branes.  It was shown that the twist fields initiate worldsheet
spectral flow from the NS$-$ spectrum on the $\DD$ system to the NS$+$
spectrum on the DD.  The tachyon potential~(\ref{nuPotential}),
derived exactly from this formalism, has all the correct properties to
describe the recombination of intersecting branes.

We have calculated the canonical kinetic term for the tachyon, but
because the twist fields intertwine with the worldsheet degrees of
freedom, the complete dependence of the action on tachyon first
derivatives cannot be obtained by our methods.  There will therefore
be $\mathcal O(\partial T)^4$ corrections to the low energy effective
action~(\ref{LEEA}), in addition to higher derivative corrections.
The study of vortices for a system of angled branes has many important
applications~\cite{Jones:2002cv,Sarangi:2002yt}, but a thorough
understanding of vortices requires higher kinetic terms.  For
instance, in order to study the lower dimensional D-branes as solitons
in the $\DD$ system it is necessary to know the action to all powers
of $\partial T$~\cite{Kraus:2000nj,Takayanagi:2000rz}.  While the
complete expression for the kinetic term requires solving an
interacting worldsheet theory, it may be possible to simply deduce the
$\partial T \to \infty$ behaviour directly from the worldsheet
boundary action.

Another issue of future interest is the inclusion of gauge fields and
scalars on the diagonal of the boundary perturbation matrix.
Transverse gauge fields may be included by covariantizing the
spacetime derivatives in the low energy effective action.  In the
twisted directions, we may insert scalar/gauge field perturbations on
the worldsheet boundary directly.  In the simplest scenario, their
inclusion would modify the boundary interaction in
~(\ref{TwistedBAction}) by adding terms of the form $\sim\Phi X^\prime
+ A \dot{X}+$(derivatives), $\Phi$, $A$ being scalar and gauge fields
in the twisted plane.  Typically, this modification makes the
worldsheet theory interacting and is more difficult to handle. Yet, it
may be interesting to apply a weak field expansion to evaluate the
resulting partition function, perhaps to resolve more of the
recombination dynamics in the twisted plane.

Finally, brane-brane separation has been suggested as an inflaton
candidate for the inflationary scenario in the early universe.  Such
an action describing lower-dimensional angled $Dp$-branes can be
easily obtained by T-dualizing the above action.  In the current
scenario of angled branes, toward the end of the inflationary epoch,
as the D$p$-branes approach each other, the tachyon mode appears.  The
tachyon rolls as the branes recombine, and the energy released goes to
some combination of tachyon matter, closed string modes, open string
modes and defects.  For the universe to enter the big bang epoch,
there are strong limits on the production of tachyon matter, closed
string modes and defects.  To study the brane dynamics and the
conditions imposed by cosmology, an effective action is desirable.  It
is this application to cosmology that motivates us to construct the
effective action for angled branes.  Some of these applications will
be discussed in a later publication.

\acknowledgments
 
Vatche Sahakian participated in some stages of this work.  His deep
and insightful comments have been invaluable.  We also thank Koji
Hashimoto, Louis Leblond and Horace Stoica for useful discussions.
This material is based upon work supported by the National Science
Foundation under Grant No.~PHY-0098631.

\appendix

\section{Superspace conventions}\label{SuAppendix}\label{section8}
 
Since there are many variations on the definitions of the worldsheet
superspace quantities, in this appendix we summaries those used in
this work.  We follow mostly the conventions of
~\cite{Polchinski:1998rr}.  The worldsheet bulk has $\N = (1,1)$
supersymmetry, which can be seen most easily with holomorphic and
anti-holomorphic superspace coordinates, $\mb z = (z, \sCoord)$,
$\bar{\mb z} = (\bar z, \bar\sCoord)$.  Define superderivatives and
supercharges
 \begin{equation}
\begin{array}[b]{rclcrcl}
   \sD{\sCoord} &=& \partial_\sCoord + \sCoord\partial\,,&\qquad&
   \sD{\bar\sCoord} &=& \partial_{\bar\sCoord} 
   + \bar\sCoord\bar\partial\,,\\
   \sC{\sCoord} &=& \partial_\sCoord - \sCoord\partial\,,
   &\qquad& \sC{\bar\sCoord} &=& \partial_{\bar\sCoord} 
   - \bar\sCoord\bar\partial\,,
 \end{array}
\end{equation}
which satisfy the relations
\begin{equation}
\begin{array}[b]{rclcrcl}
   \bigl\{\sD{\sCoord},\sD{\sCoord}\bigr\} 
   &=& -\bigl\{\sC{\sCoord},\sC{\sCoord}\bigr\} = 2\partial\,,&\qquad&
   \bigl\{\sD{\bar\sCoord},\sD{\bar\sCoord}\bigr\} 
   = -\bigl\{\sC{\bar\sCoord},\sC{\bar\sCoord}\bigr\} = 2\bar\partial\,,
\\
   \bigl\{\sD{\sCoord},\sD{\bar\sCoord}\bigr\} &=& 
   \bigl\{\sC{\sCoord},\sC{\bar\sCoord}\bigr\} = 
   \bigl\{\sD{\sCoord},\sC{\sCoord}\bigr\} = \cdots = 0\,.&\qquad& &{}&
 \end{array}
\end{equation}
The conformal field theory operator which generates holomorphic
supersymmetry transformations is the worldsheet supercurrent,
 \begin{equation}
   T_F(z) \equiv i\sqrt{\frac2{\ap}} \psi^\mu\partial X_\mu(z) \,.
 \end{equation}
A holomorphic superfield $\mathcal O(\mb z) = \mathcal O_0(z) +
\sCoord \mathcal O_1(z)$ of conformal dimension $(h,0)$ has OPEs with
$T_F$,
 \begin{equation}
   T_F(z)\mathcal O_0(0) = \frac{-\mathcal O_1(0)}z + \cdots\,,\qquad
   T_F(z)\mathcal O_1(0) = -\frac{2h \mathcal O_0(0)}{z^2}
   - \frac{\partial\mathcal O_0}z + \cdots\,. 
\label{TFOPEs}
\end{equation}
The anti-holomorphic supercurrent has similar action.  The worldsheet
bosons and fermions are grouped into a superfield,
 \begin{equation}
\label{SuperX}
   \mb X^\mu(\mb z,\bar{\mb z}) = X^\mu(z,\bar z) +
   i\sCoord\sqrt{\frac{\ap}2}\psi^\mu(z) +
   i\bar{\sCoord}\sqrt{\frac{\ap}2}\tilde\psi^\mu(\bar z)
   + \sCoord\bar\sCoord F^\mu(z,\bar z)\,.
 \end{equation}
We use the convention that for two complex Grassmanian quantities,
$\overline{\sCoord_1\sCoord_2} = \bar\sCoord_2\bar\sCoord_1$, so $\mb
X$ is a real scalar superfield. It is a simple matter to rewrite the
supersymmetric action on the worldsheet,
$$
   S_\Sigma = \frac1{2\pi\ap}\int d^2z\; d^2\sCoord\;
   \sD{\bar\sCoord}\mb X^\mu \sD{\sCoord}\mb X_\mu\,,
$$ 
and $F$ can be integrated out being auxiliary.

\subsection{Upper half plane}\label{section8.1}

On the disk represented by the upper half plane, along the boundary,
$z = \bar z \equiv y$ the 2D $\N = (1,1)$ SUSY reduces to $\N = 1$
SUSY in 1 dimension.  Thus the superspace has the boundary $\sCoord =
\pm\bar\sCoord$; choosing the ``+'' sign, defining $\BsCoord =
\sCoord$ and following~\cite{Hori:2000ic}, we see that only the linear
combination of supercharges and superderivatives which preserve the
boundary conditions are conserved,
$$
   \sC{\BsCoord} = \sC{\sCoord} + \sC{\bar\sCoord} = \partial_\BsCoord
   - \BsCoord\partial_y\,,\qquad
   \sD{\BsCoord} = \sD{\sCoord} + \sD{\bar\sCoord} = \partial_\BsCoord
   + \BsCoord\partial_y\,,
$$ On the boundary because we have $\psi = \pm\tilde\psi$ (where the
sign must match that chosen for the supercoordinate),~(\ref{SuperX})
reduces to
$$
   \mb X^\mu(\mb y) = X^\mu(y) + i\BsCoord \sqrt{2\ap} \psi^\mu(y)\,.
$$ The boundary condition on the energy momentum tensor is $T_B(z) =
\tilde T_B(\bar z)$.  This can be obtained from the methods of
~\cite{Polchinski:1998rq}, by imposing that the conserved worldsheet
current crossing the boundary is zero.  We can use this boundary
condition and that on the coordinates to represent the
anti-holomorphic sector as the lower half plane reflection of the
holomorphic sector as is usual.  Then we obtain a relation between the
Virasoro generators,
 \begin{equation}\label{UHPModes}
    L_m = \oint\limits_\Gamma \frac{dz}{2\pi i}
    (z-z_0)^{m+1}T_B(z-z_0)
    = \oint\limits_{\Gamma^\prime} \frac{d\bar z}{2\pi i}
    (\bar z - \bar z_0)^{m+1} \tilde T_B(\bar z - \bar z_0) 
    = \tilde L_m\,,
 \end{equation}
where the contour $\Gamma$ is closed in the UHP about $z_0$, and
$\Gamma^\prime$ is its reflection about the real axis.  Similar
results hold for the the worldsheet supercurrent: the boundary
condition that none of the conserved current in a worldsheet SUSY
transform flows across the boundary is $T_F(z) = \tilde T_F(\bar z)$,
which leads to
$$ G_r = \oint\limits_\Gamma \frac{dz}{2\pi i} (z-z_0)^{r+1/2}T_F(z -
z_0) = \oint\limits_{\Gamma^\prime} \frac{d\bar z}{2\pi i} (\bar z -
\bar z_0)^{r+1/2} \tilde T_F(\bar z - \bar z_0) = \tilde G_r\,.
$$ 
Note that these identifications are consistent with the holomorphic
and anti-holomorphic super-Virasoro algebras.

From these results, some general statements can be made about
operators on the boundary.  The condition $L_0 = \tilde L_0$ from
~(\ref{UHPModes}) implies that $h = \tilde h$, and the condition
$G_{-1/2} = \tilde G_{-1/2}$ implies that the holomorphic and
anti-holomorphic superpartners of a field are equal on the boundary,
for instance $\psi = \tilde\psi$ and $\partial X = \bar\partial X$.
That $h = \tilde h$ is just a statement that the boundary condition on
a $(h,\tilde h)$ field links it with a $(\tilde h, h)$ field.  Also
under the transformation of a field on the boundary to the strip by $z
= e^{-iw}$, a $(h,\tilde h)$ operator on the boundary $(w = i\tau)$
acquires a factor
$$
   \mathcal O(\tau) = 
   \left(\frac{\partial w}{\partial z}\right)^{-h}
   \left(\frac{\partial\bar w}{\partial\bar z}\right)^{-\tilde h}
   \mathcal O(z,\bar z)|_{z = \bar z} 
   = i^{\tilde h - h}e^{(h+\tilde h)\tau}
   \mathcal O(z,\bar z)|_{z = \bar z}\,.
$$ Thus, the time propagator along the strip boundary is just the
dilatation generator, $L_0 + \tilde L_0$, on the UHP, as expected.

\subsection{Unit disk}\label{section8.2}
 
It is more common in BSFT to represent the upper half plane by the
unit disk centered on the origin.  The boundary is defined by $z =
1/\bar z \equiv e^{-i\tau}$, and the 2D $\N = (1,1)$ SUSY reduces to
$\N = 1$ SUSY in 1 dimension.  As discussed in ~\cite{Itoyama:1987qz}
the complex Grassman space has the boundary $\sCoord = \mp
i\bar\sCoord/\bar z$.  Taking the $-$ sign we define the real Grassman
boundary coordinate $\BsCoord \equiv (-iz)^{-1/2}\sCoord$; the linear
combination of supercharges and superderivatives which preserve the
boundary conditions are conserved,
$$
   \sC{\BsCoord} = (-iz)^\hf\sC{\sCoord} 
   + (i\bar z)^\hf\sC{\bar\sCoord} = \partial_\BsCoord
   - \BsCoord\partial_\tau\,,\qquad
   \sD{\BsCoord} = (-iz)^\hf\sD{\sCoord} 
   + (i\bar z)^\hf\sD{\bar\sCoord} = \partial_\BsCoord
   + \BsCoord\partial_\tau\,.
$$ This is equivalent to transforming these quantities of conformal
dimension $1/2$ to the $(\sigma,\tau)$ coordinate system where $z =
e^{-\sigma-i\tau}$, and noting that on the boundary $(\sigma = 0)$,
only the $\tau$ components of the supercharge and superderivative are
conserved.  Note that the branch cut in the definitions of the
Grassman quantities in $(\tau,\BsCoord)$ coordinates ensures that
these quantities acquire a ``$-$'' sign as they go around the disk, as
is necessary.  The fermion boundary condition is $\psi = i\bar
z\tilde\psi$, and transforming the fermion to the $(\sigma,\tau)$
coordinates of the disk $\psi \to (-iz)^{-1/2}\psi$ with $\psi$ now a
real fermion,~(\ref{SuperX}) becomes on the boundary
$$
   \mb X^\mu(\mb\tau) = X^\mu(\tau) 
   + i\BsCoord \sqrt{2\ap} \psi^\mu(\tau)\,.
$$ In the literature on BSFT, there is much variation on these
conventions, but these adhere closely to those in
~\cite{Takayanagi:2000rz} (in which $\ap$ is set to 2).

The boundary condition on the energy momentum tensor is $z^2T_B(z) =
\bar z^2\tilde T_B(\bar z)$, and that on the supercurrent is $(-i)^1/2
z^{\frac32}T_F(z) + i^1/2\bar z^{\frac32}\tilde T_F(\bar z) = 0$.
This and the reflection, $z = 1/\bar z$, give the relations between
the Virasoro and supercurrent generators
 \begin{equation}
\begin{array}[b]{rclcrcl}
    L_m &=& \displaystyle\oint\limits_\Gamma \frac{dz}{2\pi i}
    (z - z_0)^{m+1}T_B(z - z_0)
    &\qquad& G_r &=& \displaystyle\oint\limits_\Gamma \frac{dz}{2\pi i}
    (z - z_0)^{r+1/2}T_F(z - z_0)
    \\[-2pt]
    &=& \displaystyle-\oint\limits_{\Gamma^\prime} \frac{d\bar z}{2\pi i}
    (\bar z - \bar z_0)^{-m+1}\tilde T_B(\bar z - \bar z_0)
    &\qquad&  &=& \displaystyle i\oint\limits_{\Gamma^\prime} \frac{d\bar z}{2\pi i}
    (\bar z - \bar z_0)^{-r+1/2}\tilde T_F(\bar z - \bar z_0)
\\[-2pt]
    &=& -\tilde L_{-m}\,,&\qquad&
    &=& i\tilde G_{-r}\,,\label{DiscModes}
 \end{array}
\end{equation}
 
\EPSFIGURE{DiscContours.eps}{\label{DiscContours}
   Disk contours.}
\noindent 
 where the appropriate contours are shown in
 figure~\ref{DiscContours}.  These identifications are again
 consistent with the holomorphic and anti-holomorphic super-Virasoro
 algebras, and they imply general statements about boundary operators.
 The condition $L_0 = -\tilde L_0$ from~(\ref{DiscModes}) implies that
 $h = -\tilde h$, which is the statement that the boundary condition
 on a $(h,\tilde h)$ conformal field identifies it with a $(-\tilde h,
 -h)$ conformal field; such an example is $\psi$ with $(h,\tilde h) =
 (1/2,0)$, and the boundary condition identifies this with $i\bar
 z\tilde\psi$ which scales like $(0,-1/2 )$.  The condition $G_{-1/2}
 = i\tilde G_{1/2}$ links the holomorphic SUSY transformation of an
 operator with $i\bar z$ times its anti-holomorphic SUSY
 transformation.  The fermion boundary condition, $\psi = i\bar
 z\tilde\psi$, is again an example.  The transformation to polar
 coordinates parameterizing the disk, $z = e^{-\sigma-i\tau}$, sees a
 $(h,\tilde h)$ operator on the boundary $(\sigma = 0)$ acquire the
 factor
$$
   \mathcal O(\tau) = 
   \left(\frac{\partial w}{\partial z}\right)^{-h}
   \left(\frac{\partial\bar w}{\partial\bar z}\right)^{-\tilde h}
   \mathcal O(z,\bar z)|_{z = 1/\bar z} 
   = i^{\tilde h - h}e^{-i(h-\tilde h)\tau}
   \mathcal O(z,\bar z)|_{z = 1/\bar z}\,.
$$ This suggests that the time propagator along the boundary of the
disk is just the rotation generator of the plane, $L_0 - \tilde L_0$.

On the boundary, both the holomorphic and anti-holomorphic SUSY
supercurrents contribute to the SUSY transform of an operator because
the boundary condition links the left and right SUSY transformations.
The SUSY transformation of a boundary operator is then given by an
integral of the OPE of that operator with the supercurrent, where the
contour goes half way about the pole location.  The result is that for
a boundary operator,
 \begin{eqnarray}
   T_F(\tau) &=& (-iz)^{3/2}T_F(z) 
   + (i\bar z)^{3/2}\tilde T_F(\bar z) 
   = i\sqrt{\frac2{\ap}} \psi^\mu\dot X_\mu(\tau)\,,
\nonumber\\[-2pt]
   T_F(\tau)\mathcal O_0(0) &=& \frac{-\mathcal O_1(0)}{\tau}
   + \cdots\,.
 \end{eqnarray}
A twist superfield can now be constructed using these relations.  The
boundary field of lowest conformal dimension which keeps $T_F$ single
valued while twisting the bosons and fermions is $\twist_0^\pm \equiv
(\sigma s)^\pm$.  The upper component is given by
 \begin{eqnarray}
&&   T_F(\tau)\twist_0^\pm(0) = -\frac{\twist_1^\pm(0)}\tau 
   + \cdots\,,\\[-2pt] 
&&   \Longrightarrow \mb\twist^\pm(\mb\tau) = 
   (\sigma s)^\pm(\tau) 
   - \frac i{\sqrt{2\ap}}\BsCoord(\mu u)^\pm(\tau)\,.
 \end{eqnarray}
We have used the OPEs~(\ref{BTwistOPEs}) and~(\ref{FTwistOPEs}).

\pagebreak[3]

\end{document}